\documentclass[showpacs,aps,twocolumn]{revtex4}
\usepackage{bm}
\usepackage{amsmath}
\usepackage{graphicx}
\usepackage{subfigure}
\usepackage[usenames,dvipsnames]{color}
\definecolor{darkblue}{RGB}{0,0,196}
\usepackage[colorlinks=true,linkcolor=darkblue,citecolor=darkblue,urlcolor=darkblue]{hyperref}
\usepackage{setspace}
\usepackage{hyperref}
\usepackage{xcolor}
\usepackage{footmisc}
\usepackage[makeroom]{cancel}
\usepackage{comment}
\usepackage{lineno}
\def\be{\begin{equation}}
\def\ee{\end{equation}}
\def\ba{\begin{eqnarray}}
\def\ea{\end{eqnarray}}
\usepackage{graphicx}
\usepackage{amsmath,bbm}
\usepackage{amssymb,bm}
\usepackage{yfonts}

\begin{document}
\title{Event topology and constituent-quark scaling of elliptic flow in heavy-ion collisions at the Large Hadron Collider using a multiphase transport model}

\author{Neelkamal Mallick$^1$}
\author{Sushanta Tripathy$^2$}
\author{Raghunath Sahoo$^{1,3}$\footnote{Corresponding author: $Raghunath.Sahoo@cern.ch$}}

\affiliation{$^1$Department of Physics, Indian Institute of Technology Indore, Simrol, Indore 453552, India}
\affiliation{$^2$INFN - sezione di Bologna, via Irnerio 46, 40126 Bologna BO, Italy}
\affiliation{$^{3}$CERN, CH 1211, Geneva 23, Switzerland}

\begin{abstract}
\noindent
Transverse spherocity is an event shape observable, which separates the events based on their geometrical shapes. In this work, we use transverse spherocity to study the identified light flavor production in heavy-ion collisions using A Multi-Phase Transport (AMPT) model. We obtain the elliptic flow coefficients for pions, kaons and protons in Pb+Pb collisions at $\sqrt{s_{\rm{NN}}} = 5.02$~TeV as a function of transverse spherocity and collision centrality. Also, we study the number of constituent-quark (NCQ) scaling of elliptic flow which interprets the dominance of the quark degrees of freedom at the early stages of the collision. We observe a clear dependence of the elliptic flow for identified particles on transverse spherocity. It is found that the NCQ-scaling is strongly violated in events with low transverse spherocity compared to transverse spherocity-integrated events, confirming the fragmentation-based hadronization mechanism for high-momentum partons involved in the dynamics of jetty-like events.

%\pacs{13.85.Ni,12.38.Mh, 25.75.Nq, 25.75.Dw}
\end{abstract}
\date{\today}
\maketitle 

\section{Introduction}
\label{intro}
The ultra-relativistic heavy-ion collisions at particle accelerators like the Large Hadron Collider (LHC) at CERN, Switzerland, and Relativistic Heavy-Ion Collider (RHIC) at BNL, USA, provide signatures of the possible formation of a deconfined state of quarks and gluons, known as Quark-Gluon Plasma (QGP). Although, one does not observe any direct evidence of possible QGP formation due to its very short lifetime, instead several indirect signatures such as direct photon measurements, jet-quenching, elliptic flow ($v_2$), strangeness enhancement, quarkonia suppression etc. suggest that the formation of QGP is highly probable in such collisions. Due to the unprecedented energy scale of a few TeV, some of the QGP signatures~\cite{ALICE:2017jyt,Khachatryan:2016txc} are also observed in the smallest collision systems (i.e. proton-proton collisions) at the LHC. To understand the underlying dynamics of small collision systems, an event shape observable, transverse spherocity, has been introduced recently~\cite{Cuautle:2014yda,Cuautle:2015kra,Salam:2009jx,Bencedi:2018ctm,Banfi:2010xy} at the LHC energies. From these studies, it was observed that transverse spherocity has the potential to separate the events based on their geometrical shapes, i.e. isotropic and jetty events ~\cite{Tripathy:2019blo,Khuntia:2018qox}. After its successful implementation in small collision systems, transverse spherocity can be used as a tool for heavy-ion collisions to differentiate events based on their event topology, even though the environment is much more complex. It might reveal new results from heavy-ion collisions where the production of a QGP medium is already established. In one of our recent publications~\cite{Mallick:2021rsd}, we have explicitly used transverse spherocity in heavy-ion collisions for the first time to study the azimuthal anisotropy of unidentified charged particles as a function of transverse spherocity in A Multi-Phase Transport (AMPT) model. We observe a strong anti-correlation of transverse spherocity with the ellipticity of the events in heavy-ion collisions. It was found that high transverse spherocity events have nearly zero $v_2$ while low transverse spherocity events contribute significantly to the $v_2$. 

$v_2$ is generally caused by the initial spatial anisotropy in the system produced in a non-central collision. It plays a significant role to understand the collective motion and bulk property of the QGP. $v_2$ is defined as the second-order Fourier component of the particle azimuthal distribution, which provides information on the transport properties of the created medium in heavy-ion collisions~\cite{v2}. The comparison of $v_2$ measurements to hydrodynamic calculations reveal that the elliptic flow is built up mainly during the early partonic stages of QGP and it is governed by the evolution of  QGP~\cite{ Abelev:2014pua,Arsene:2004fa,Adcox:2004mh,Back:2004je,Adams:2005dq,ATLAS:2011ah,ATLAS:2012at,CMS:2012xss,CMS:2012zex}. However, the hadronic rescattering in the hadronic phase could also contribute to the evolution of elliptic flow~\cite{Hirano:2005xf}. Thus, to understand the interplay of partonic versus hadronic phase in the evolution of elliptic flow, it is important to study the elliptic flow of different identified particles.

\begin{figure*}[ht]
\includegraphics[scale=0.8]{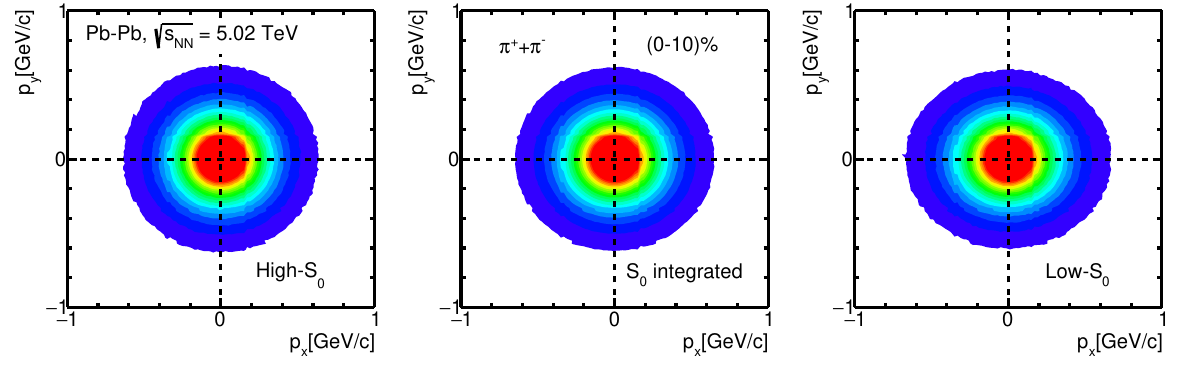}
\includegraphics[scale=0.8]{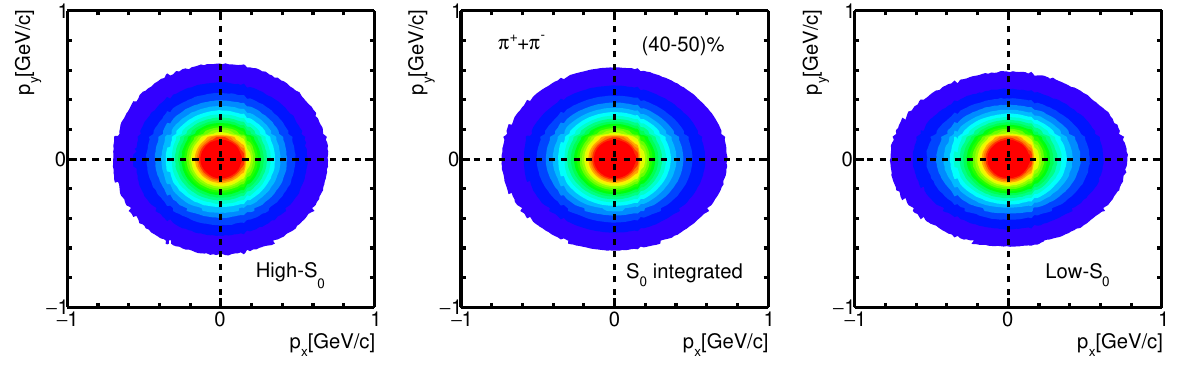}
\caption[]{(Color Online) Transverse momentum space correlation ($p_{\rm y}$ vs. $p_{\rm x}$) for $\pi^++\pi^-$ in different spherocity classes for (0-10)\% (top) and (40-50)\% (bottom) central Pb+Pb collisions at $\sqrt{s_{\rm{NN}}} = 5.02$~TeV in AMPT model.}
\label{fig1}
\end{figure*}

In addition, RHIC and LHC experimental results found that in the intermediate transverse momentum range, the $v_2$ of baryons is larger than that of mesons~\cite{Abelev:2014pua}. Ref.~\cite{Voloshin:2002wa} suggests that this phenomenon could be explained by the development of hydrodynamic flow in the partonic phase and then quarks coalesce into hadrons during the hadronization. This mechanism leads to the number of constituent quarks (NCQ) scaling, a hierarchy observed in the values of $v_2$ as a function of transverse momentum, where the flow might no longer be dominant and this may compete with the contribution from jet fragmentation. This scaling property has been studied extensively in experiments~\cite{Adams:2003am,Adler:2003kt,Abelev:2014pua, v2_ALICE2} and in theory~\cite{Molnar:2003ff,Greco:2003mm,Fries:2003kq}. Some of these results suggest that the NCQ scaling of $v_{2}$ at the LHC energies is violated~\cite{Abelev:2014pua, v2_ALICE2} for intermediate or high transverse momenta for heavy-ion collisions. In RHIC experimental results for $\sqrt{s_{\rm{NN}}} = 200$~GeV~\cite{Adams:2003am,Adler:2003kt}, NCQ scaling behavior was observed for identified particles. As one goes down further in collision energy, the NCQ-scaling is reported to be violated because of the bare energy density required to form a partonic phase and also because of baryon transport to the mid-rapidity owing to baryon stopping \cite{Dunlop:2011cf}.  For Au+Au collisions at the lower RHIC energy of $\sqrt{s_{\rm{NN}}} = 12.3$~GeV, AMPT with string melting shows a nice NCQ-scaling of identified 
particle $v_2$ due to the coalescence of partons in the hadronization. However, the use of the default AMPT with dominant hadronic interactions seems to break the scaling, as hadrons are formed directly from the fragmentation of excited strings \cite{Tian:2009wg}. This makes a 
nice case of a correlation of NCQ-scaling of $v_2$ and the formation of a partonic phase of matter.
AMPT model with string melting has shown that NCQ-scaling of $v_2$ holds good at the top RHIC energy for Au+Au collisions \cite{Singha:2016aim}. The breaking of NCQ-scaling for Pb+Pb collisions at $\sqrt{s_{\rm{NN}}} = 2.76$~TeV with varying partonic scattering
cross-sections and lifetime of the hadronic cascade are presumably understood to be because of the high phase-space density of constituent quarks.
Further, smaller collision systems like Si+Si at the same collision energy showing NCQ-scaling of  $v_2$ adds to the above understanding \cite{Singha:2016aim}. In the ambit of a hydrodynamics-coalescence-fragmentation model (low-$p_{\rm T}$ hadrons undergo hydrodynamic expansion with high-$p_{\rm T}$ hadrons from the fragmentation of quenched jets), quark coalescence explains the approximate NCQ-scaling of $v_2$ observed in experimental data for the high-multiplicity p+Pb collisions at $\sqrt{s_{\rm{NN}}} = 5.02$~TeV \cite{Zhao:2019ehg}.

In this work, we obtain the transverse momentum spectra, two-particle correlations and elliptic flow of identified particles as a function of transverse spherocity in Pb+Pb collisions at $\sqrt{s_{\rm{NN}}} = 5.02$~TeV and Au+Au collisions at $\sqrt{s_{\rm{NN}}}$ = 200 GeV using AMPT model. We intend to probe the scaling properties of azimuthal anisotropy using different transverse spherocity classes in heavy-ion collisions for the first time. Thanks to the strong anti-correlation of transverse spherocity with elliptic flow~\cite{Mallick:2021rsd}, the NCQ scaling can be studied for events with different geometrical shapes (events dominated by jets versus events rich in soft particles). A multi-differential study with transverse spherocity and collision centrality can help to understand the interplay of jet fragmentation and quark coalescence mechanisms. Thus, this study has the potential to tune different phenomenological models.  We also believe that the results would motivate experimentalists to pursue such a novel method in experiments at RHIC and LHC. In fact, in a recent study, it has been shown how machine learning tools can be implemented in experiments to obtain transverse spherocity~\cite{Mallick:2021wop}. The present study based on transverse spherocity has the potential to complement the event shape engineering strategy typically implemented in heavy-ion collisions such as flow vector~\cite{Schukraft:2012ah,Aad:2015lwa,Poskanzer:1998yz}.

The paper is organized as follows. We begin with a brief introduction and motivation for the study in Section~\ref{intro}. In Section~\ref{section2}, the detailed analysis methodology along with a brief description of AMPT is given. Section~\ref{section3} discusses about the results and finally they are summarized in Section~\ref{section4}.

\section{Event Generation and Analysis Methodology}
\label{section2}

In this section, we begin with a brief introduction to the AMPT model and then we proceed with defining transverse spherocity and elliptic flow.

\subsection{A Multi-Phase Transport (AMPT) Model}
\label{formalism}
AMPT model contains four components namely, initialization, parton transport, hadronization mechanism, and hadron transport~\cite{AMPT2}.  The initialization of the collisions is performed using HIJING~\cite{ampthijing}, where the differential cross-section of the produced mini-jets in $pp$ collisions is calculated. Then, the produced partons calculated in $pp$ collisions are converted into $p$-A and heavy-ion collisions by parametrized shadowing function and nuclear overlap function with the inbuilt Glauber model. The initial low-momentum partons are produced from parametrized colored string fragmentation mechanisms and they are separated from high momentum partons by a momentum cut-off. The produced partons are propagated into the parton transport part via Zhang’s Parton Cascade (ZPC) model~\cite{amptzpc}. In the String Melting version of AMPT, the melting of colored strings into low momentum partons takes place at the start of the ZPC. It is calculated using Lund FRITIOF model of HIJING. The resulting partons undergo multiple scatterings which take place when any two partons are within a distance of the minimum approach. In AMPT-SM, the transported partons are finally hadronized using spatial coalescence mechanism~\cite{Lin:2001zk,He:2017tla}. The produced hadrons further undergo final evolution in a relativistic transport mechanism~\cite{amptart1, amptart2} via meson-meson, meson-baryon, and baryon-baryon interactions. There is also a default version of AMPT, where instead of coalescing the partons, a fragmentation mechanism using Lund fragmentation parameters $a$ and $b$ are used for hadronizing the transported partons. However, the particle flow and spectra at the mid-$p_{\rm T}$ regions are well explained by the quark coalescence mechanism for hadronization~\cite{Greco:2003mm,ampthadron2,Fries:2003kq}. Thus, we have used the string melting mode of AMPT (AMPT version 2.26t9b) for all of our calculations in the current work. The AMPT settings in the current work, are the same as reported in Refs.~\cite{Tripathy:2018bib, Mallick:2021rsd}. One should note that, to exactly match the experimental data, one can vary the tunes of the AMPT model, which is currently out of the scope of this manuscript. For the input of impact parameter values for different centralities in Pb+Pb and Au+Au collisions, we have used Ref.~\cite{Loizides:2017ack}.

\subsection{Transverse Spherocity}

%\begin{figure}[ht]
%\includegraphics[scale=0.4]{figures/Sphero.pdf}
%\caption[]{(Color Online) Spherocity distributions for (0-10)\%, (40-50)\% and (60-70)\% centrality classes in Pb+Pb collisions at $\sqrt{s_{\rm NN}}=5.02$ TeV in AMPT model.}
%\label{sp_common}
%\end{figure}

Transverse spherocity is an event property which is defined for a unit vector $\hat{n} (n_{T},0)$ that minimizes the ratio~\cite{Cuautle:2014yda, Cuautle:2015kra}:
\begin{eqnarray}
S_{0} = \frac{\pi^{2}}{4} \bigg(\frac{\Sigma_{i}~|\vec p_{T_{i}}\times\hat{n}|}{\Sigma_{i}~p_{T_{i}}}\bigg)^{2}.
\label{eq1}
\end{eqnarray}

By restricting it to the transverse plane, transverse spherocity becomes infrared and collinear safe~\cite{Salam:2009jx}. By construction, the extreme limits of transverse spherocity are related to specific configurations of events in the transverse plane i.e. jetty and isotropic. The value of transverse spherocity ranges from 0 to 1, which is ensured by multiplying the normalization constant $\pi{^2}/4$ in Eq.~\ref{eq1}. Here onwards, for the sake of simplicity the transverse spherocity is referred to as spherocity. To disentangle the low-$S_0$ and high-$S_0$ events from the spherocity-integrated events, we have applied spherocity cuts on the generated events. The spherocity distributions are selected in the pseudo-rapidity range of $|\eta|<0.8$ with a minimum constraint of 5 charged particles with $p_{\rm{T}}>$~0.15~GeV/$c$ to recreate conditions similar to ALICE experiment at the LHC~\cite{ALICE:2019dfi}. The low-$S_0$ events are those events having spherocity values in the lowest 20\% and the high-$S_0$ events are those occupying the highest 20\% in the spherocity distribution. The spherocity cuts are shown in Table~\ref{tab:1}.

\begin{table}[ht!]
\begin{center}
\caption{Lowest 20 \% (low-$S_0$) and highest 20\% (high-$S_0$) cuts on spherocity distribution for Pb+Pb collisions at $\sqrt{s_{\rm NN}} = 5.02$ TeV.} 
%For Xe-Xe collisions in 0-10 \% and 10-20 \%, we do not have significant statistics to obtain the spherocity distribution. We shall update when we have decent statistics.}
\label{tab:1}
\begin{tabular}{ |p{2.5cm}|p{1.5cm}|p{1.5cm}|p{1.5cm}|}
\hline
 Centrality (\%) & low-$S_0$ & high-$S_0$\\
\hline
0-10		  & 0-0.88025   &	0.95285-1 \\
40-50	      & 0-0.71475	  &	0.86365-1 \\
60-70         & 0-0.70775	  &	0.87355-1 \\	
 \hline
 \end{tabular}
 \end{center}
\end{table}

Figure~\ref{fig1} shows the transverse momentum space correlation ($p_{\rm y}$ vs. $p_{\rm x}$) for $\pi^++\pi^-$ in different spherocity classes for (0-10)\% (top) and (40-50)\% (bottom) central Pb+Pb collisions at $\sqrt{s_{\rm{NN}}} = 5.02$~TeV in AMPT model. The spherocity integrated events for (40-50)\% centrality class clearly show anisotropy in the momentum space which is linked to the initial state geometrical anisotropy. However, such anisotropy is not observed in the (0-10)\% centrality class. This behavior is understood as semi-central collisions develop an almond shape of the nuclear overlap region in the initial stage while for most central collisions the nuclear overlap region is nearly circular. When one studies the transverse momentum space correlation in different spherocity classes, the low-$S_0$ events show higher anisotropy compared to spherocity-integrated events while high-$S_0$ events show nearly zero anisotropy.  This is an indication that spherocity can be a useful tool for event classification to study particle production in heavy-ion collisions.
 
\begin{figure}[ht!]
\begin{center}
\includegraphics[scale=0.45]{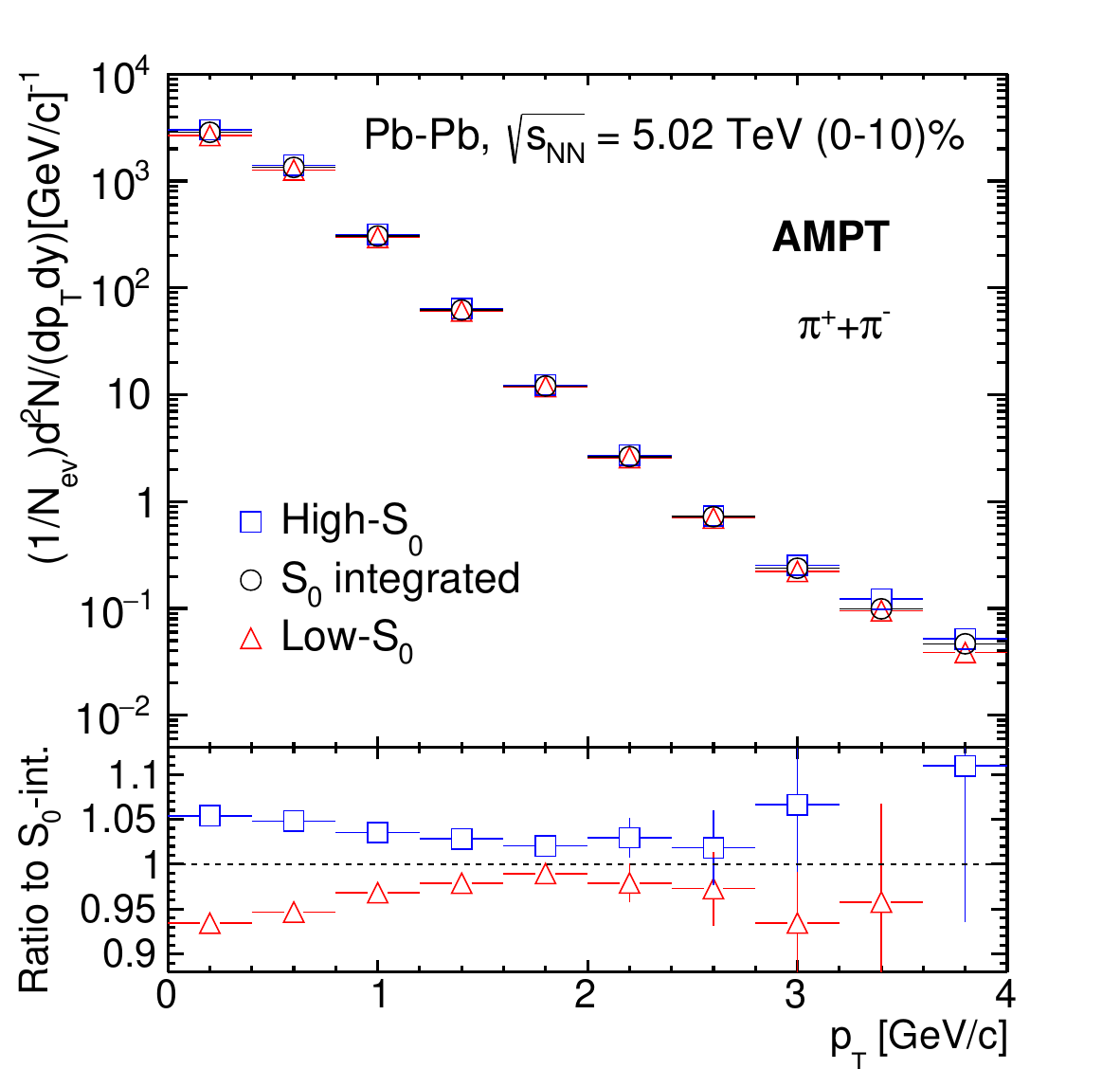}
\caption[width=18cm]{(Color Online) Top plot: $p_{\rm T}$-spectra for pions in (0-10)\% central Pb+Pb collisions for high-$S_{0}$, $S_0$-integrated and low-$S_{0}$ classes. Bottom plot: Ratio of $p_{\rm T}$-spectra for high-$S_{0}$ and low-$S_{0}$ events to the $S_0$-integrated events. }
\label{fig2}
\end{center}
\end{figure}

\subsection{Elliptic Flow}
In non-central heavy-ion collisions, the initial spatial anisotropies of the nuclear overlap region get converted to the final state momentum space azimuthal anisotropy of the produced hadrons due to the strong differential pressure gradients of the QGP medium. The observed non-zero finite azimuthal anisotropy (evident from Fig.~\ref{fig1}) can be expressed as a Fourier series in the azimuthal angle, $\phi$:
\begin{eqnarray}
E\frac{d^3N}{dp^3}=\frac{d^2N}{2\pi p_{\rm T}dp_{\rm T}dy}\bigg(1+2\sum_{n=1}^\infty v_n \cos[n(\phi -\psi_n)]\bigg)\,.\nonumber\\
\label{eq2}
\end{eqnarray}
Here, $v_n$ is the magnitude of anisotropic flow of different order $n$, and $\psi_n$ is the n$^{\text{th}}$ harmonic event plane angle~\cite{v2}. $v_n$ is usually studied as a function of transverse momentum ($p_{\rm T}$) and pseudorapidity ($\eta$).

\begin{figure}[ht!]
\begin{center}
\includegraphics[scale=0.38]{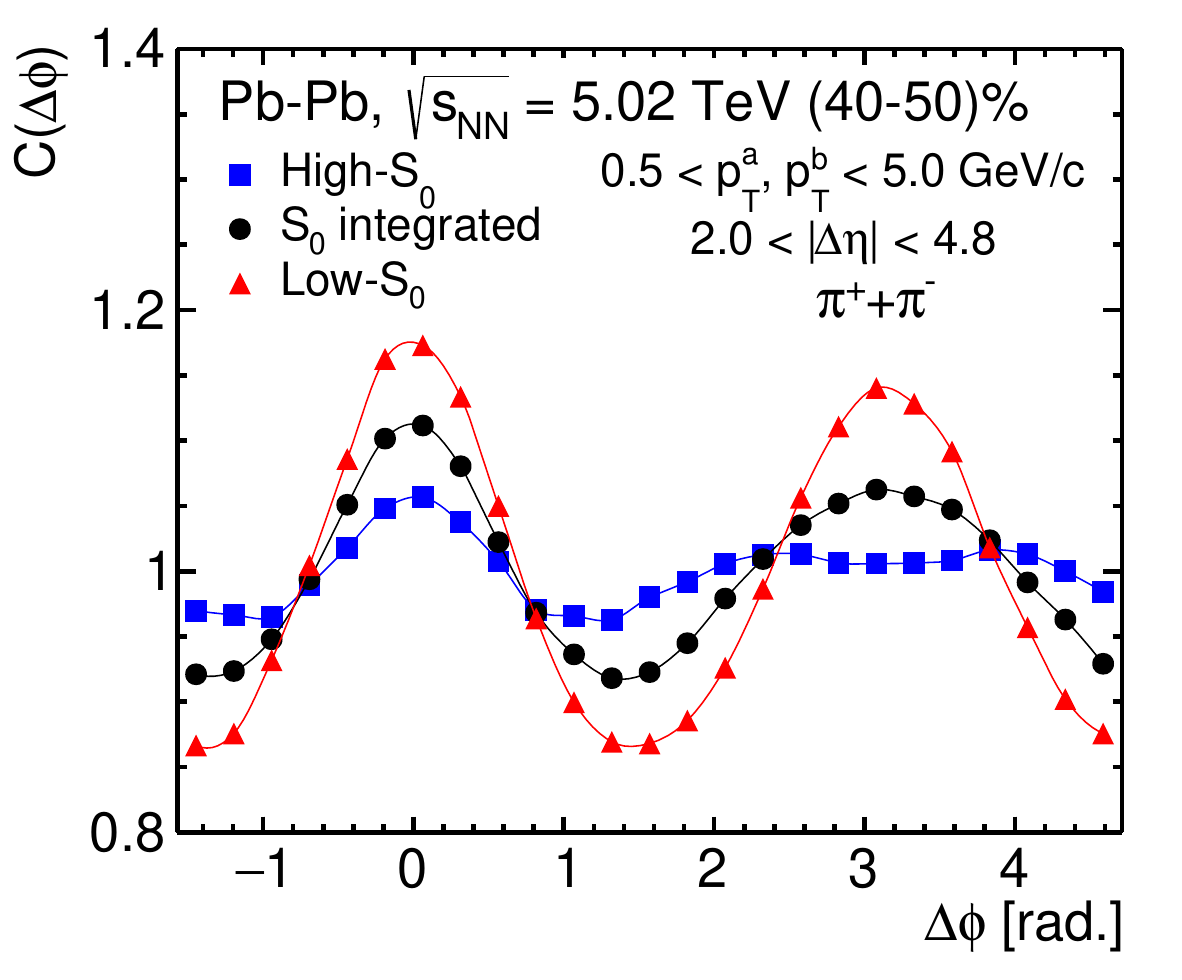}
\caption[width=18cm]{(Color Online) One dimensional azimuthal correlation of pions for low-$S_0$ (red triangles), high-$S_0$ (blue squares) and spherocity-integrated (black circles) events in 40-50\% central Pb+Pb collisions at $\sqrt{s_{\rm NN}} = 5.02$ TeV using AMPT model. }
\label{fig3}
\end{center}
\end{figure}

\begin{figure}[ht!]
\begin{center}
\includegraphics[scale=0.4]{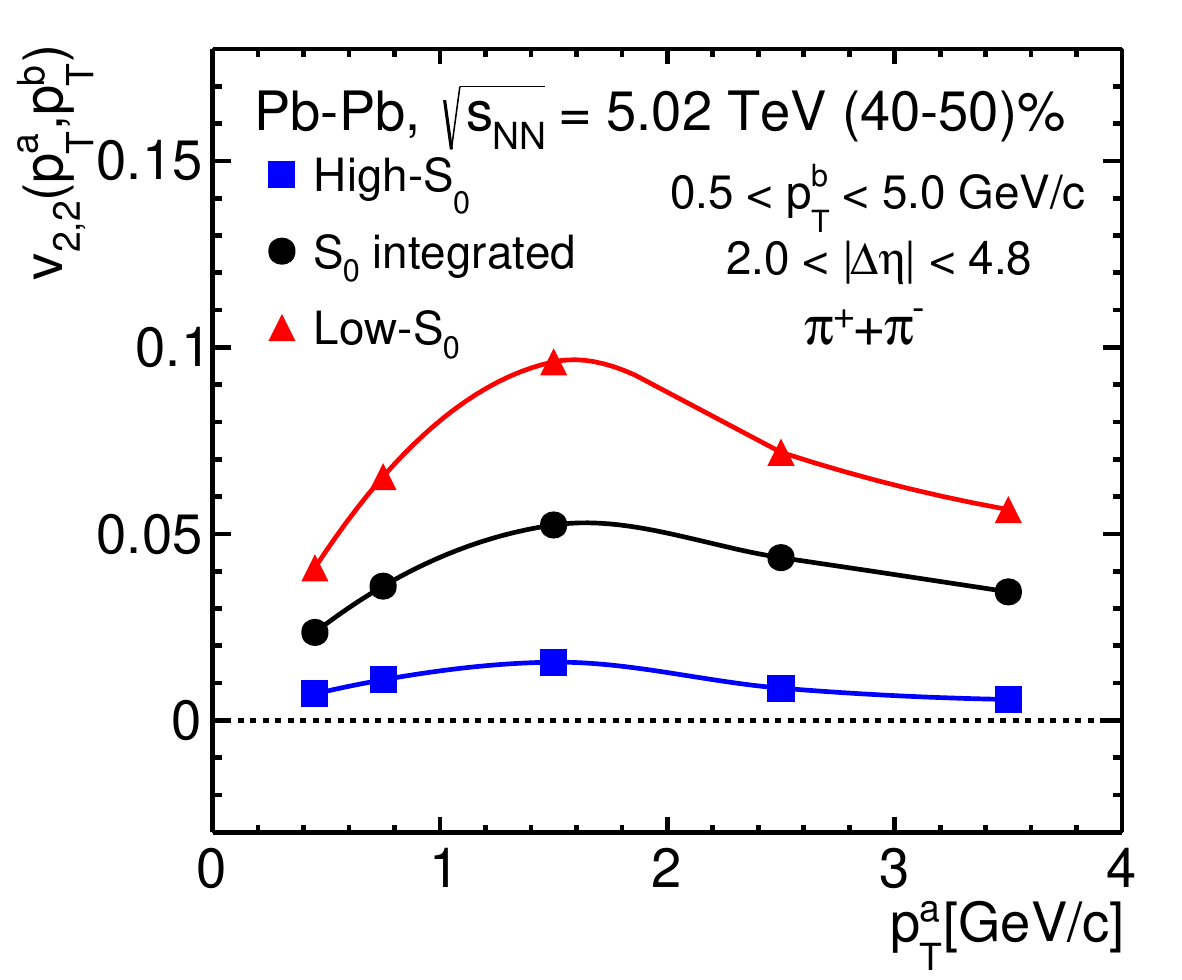}
\caption[width=18cm]{(Color Online) Two particle elliptic flow coefficients of pions for low-$S_0$ (red triangles), high-$S_0$ (blue squares) and spherocity-integrated (black circles) events in 40-50\% central Pb+Pb collisions at $\sqrt{s_{\rm NN}} = 5.02$ TeV using AMPT model. }
\label{fig4}
\end{center}
\end{figure}

\begin{figure}[ht!]
\begin{center}
\includegraphics[scale=0.4]{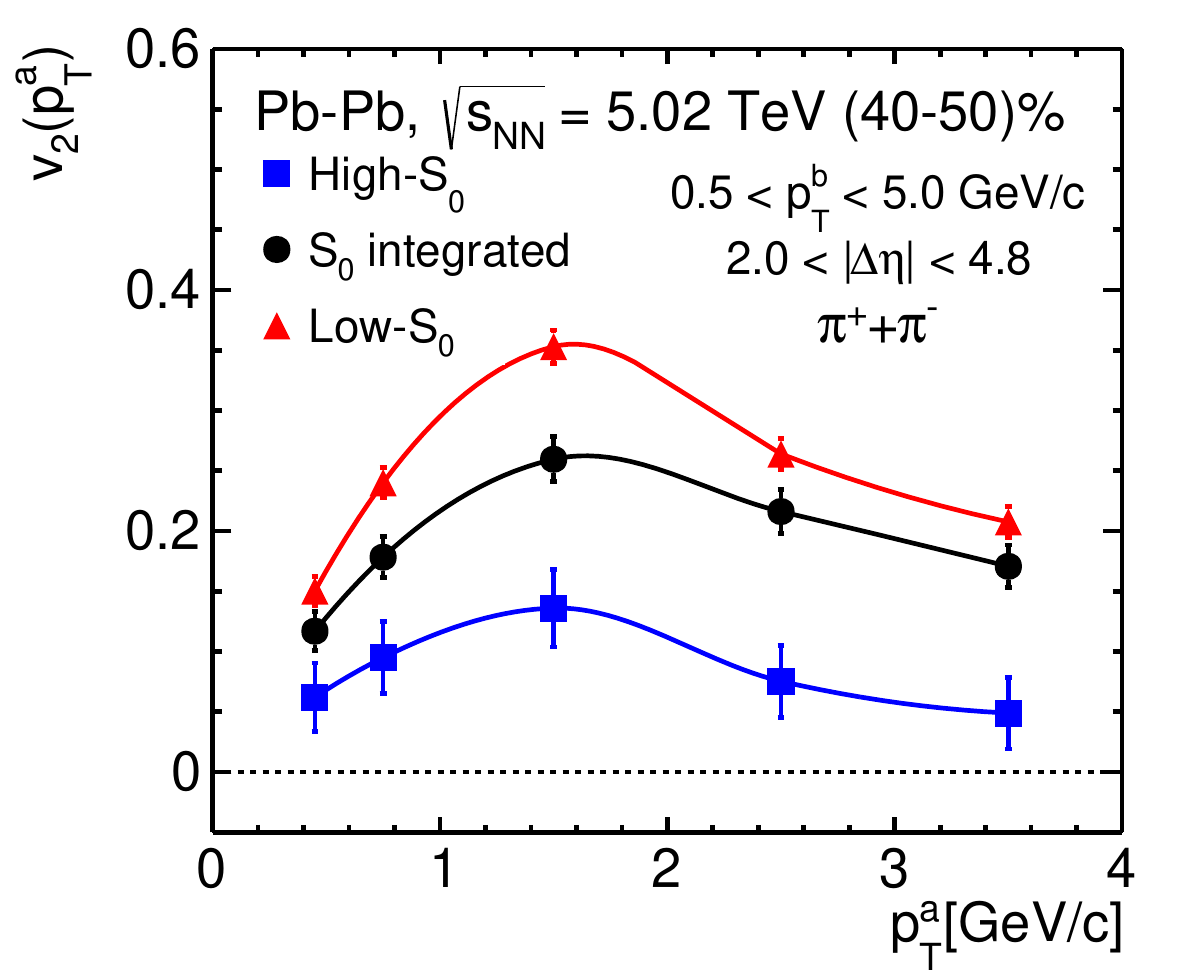}
\caption[width=18cm]{(Color Online) Single particle elliptic flow coefficients of pions for low-$S_0$ (red triangles), high-$S_0$ (blue squares) and spherocity-integrated (black circles) events in 40-50\% central Pb+Pb collisions at $\sqrt{s_{\rm NN}} = 5.02$ TeV using AMPT model. }
\label{fig5}
\end{center}
\end{figure}

%In the current work, elliptic flow is calculated with respect to the reaction plane by taking $\psi_{n}$ = 0. This implies event plane coincides with the reaction plane. Although it is non-trivial in experiments but AMPT provides the freedom to exactly define the event plane for a collision. 

%Obtaining the event plane angle is nearly impossible in real experiments. Thus, to compare with experimental data, we use two-particle correlation analysis to calculate the elliptic flow as obtained in experiments~\cite{}. The two-particle correlation method has an added advantage as by construction with a proper pseudo-rapidity cut, it would remove the residual non-flow effects in the elliptic flow. We follow a similar procedure of obtaining azimuthal anisotropy as in the LHC experiments~\cite{}, which is discussed as follows.

\begin{figure*}[ht!]
\begin{center}
\includegraphics[scale=0.29]{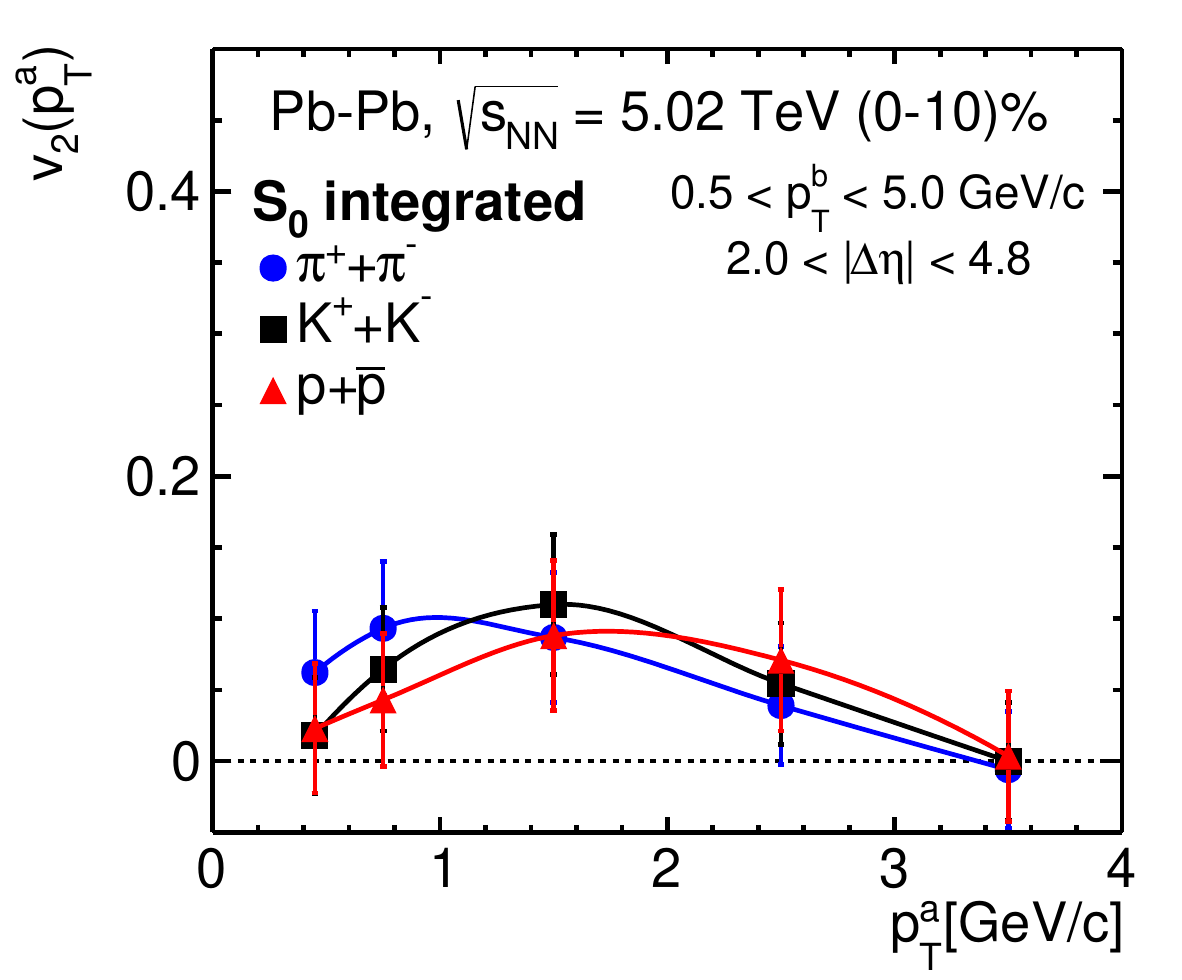}
\includegraphics[scale=0.29]{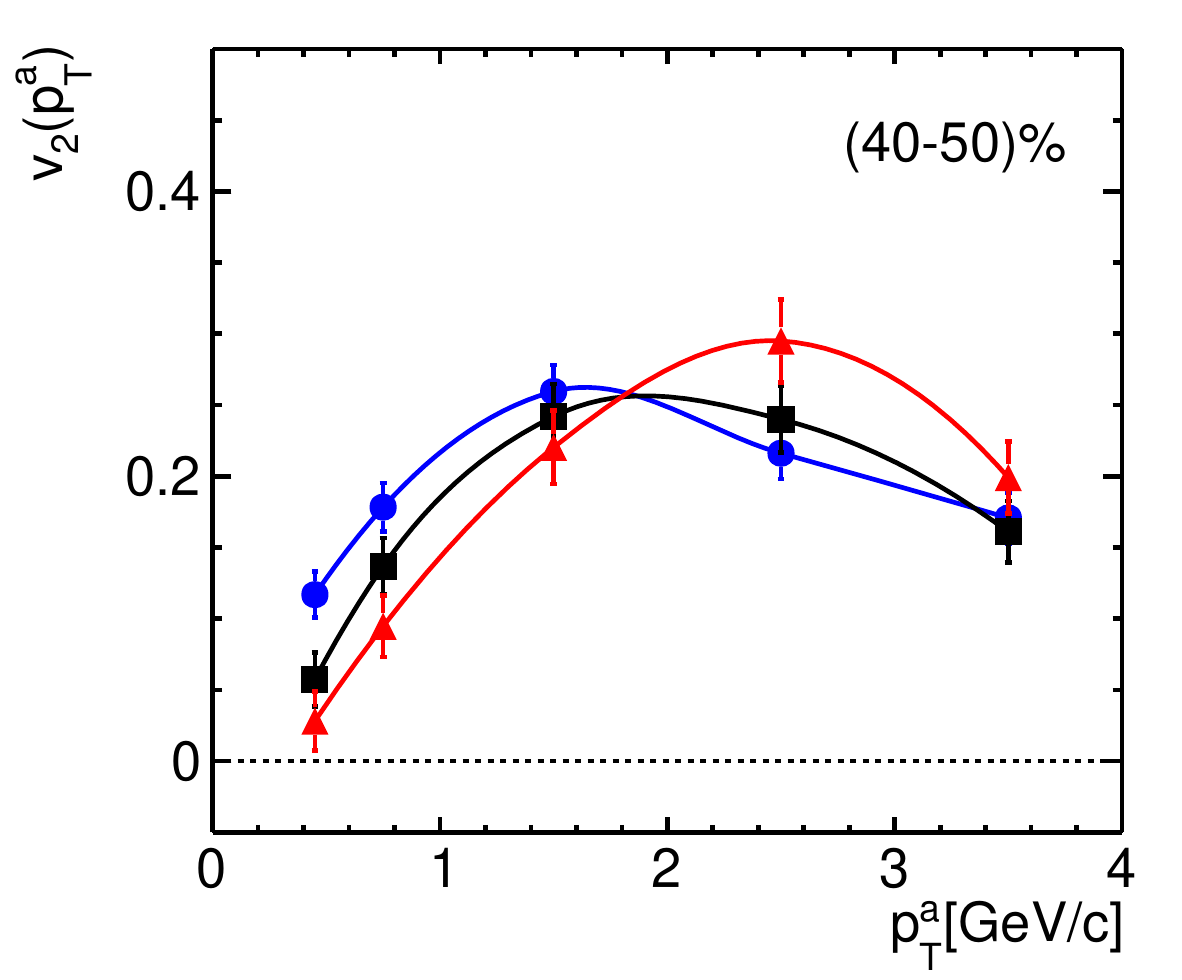}
\includegraphics[scale=0.29]{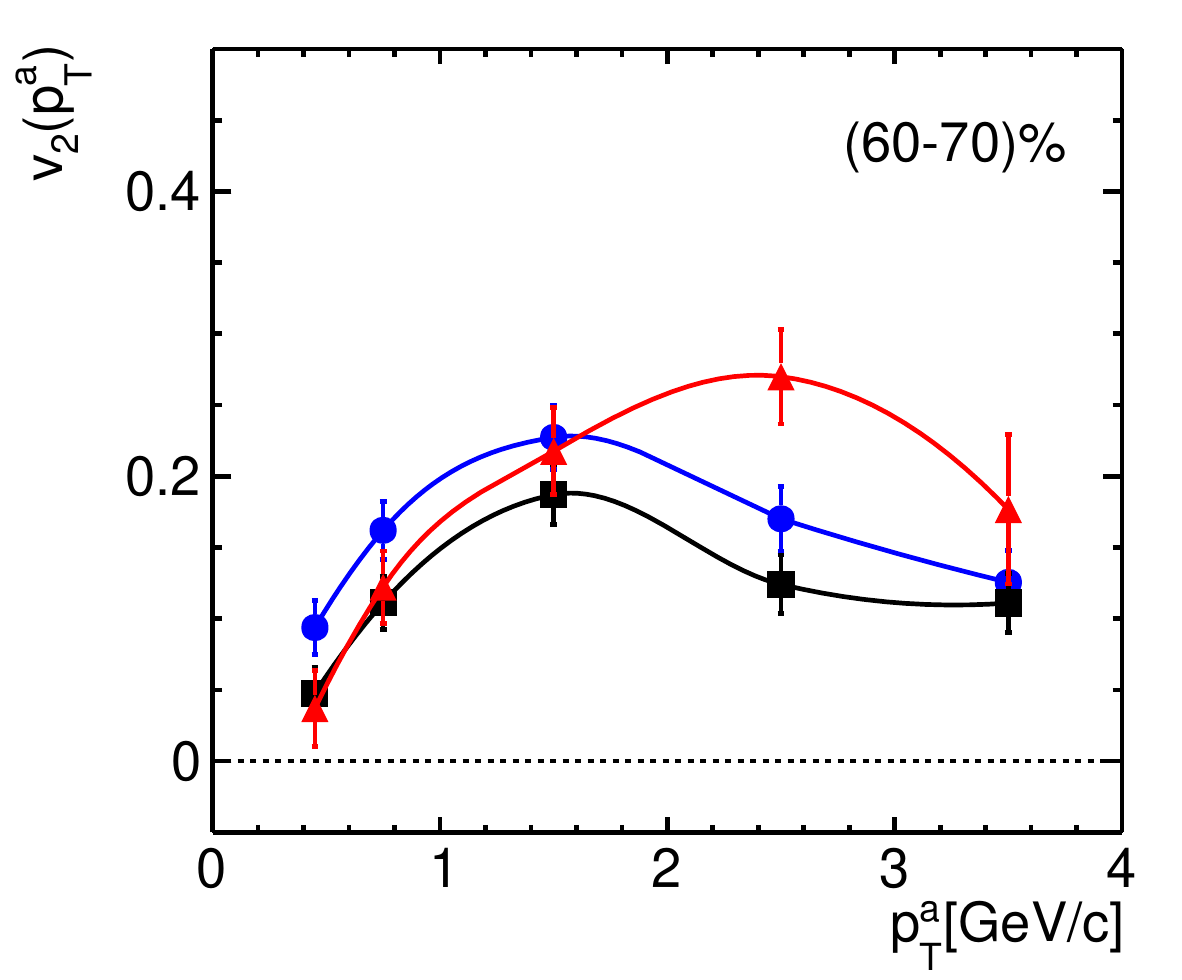}
\includegraphics[scale=0.29]{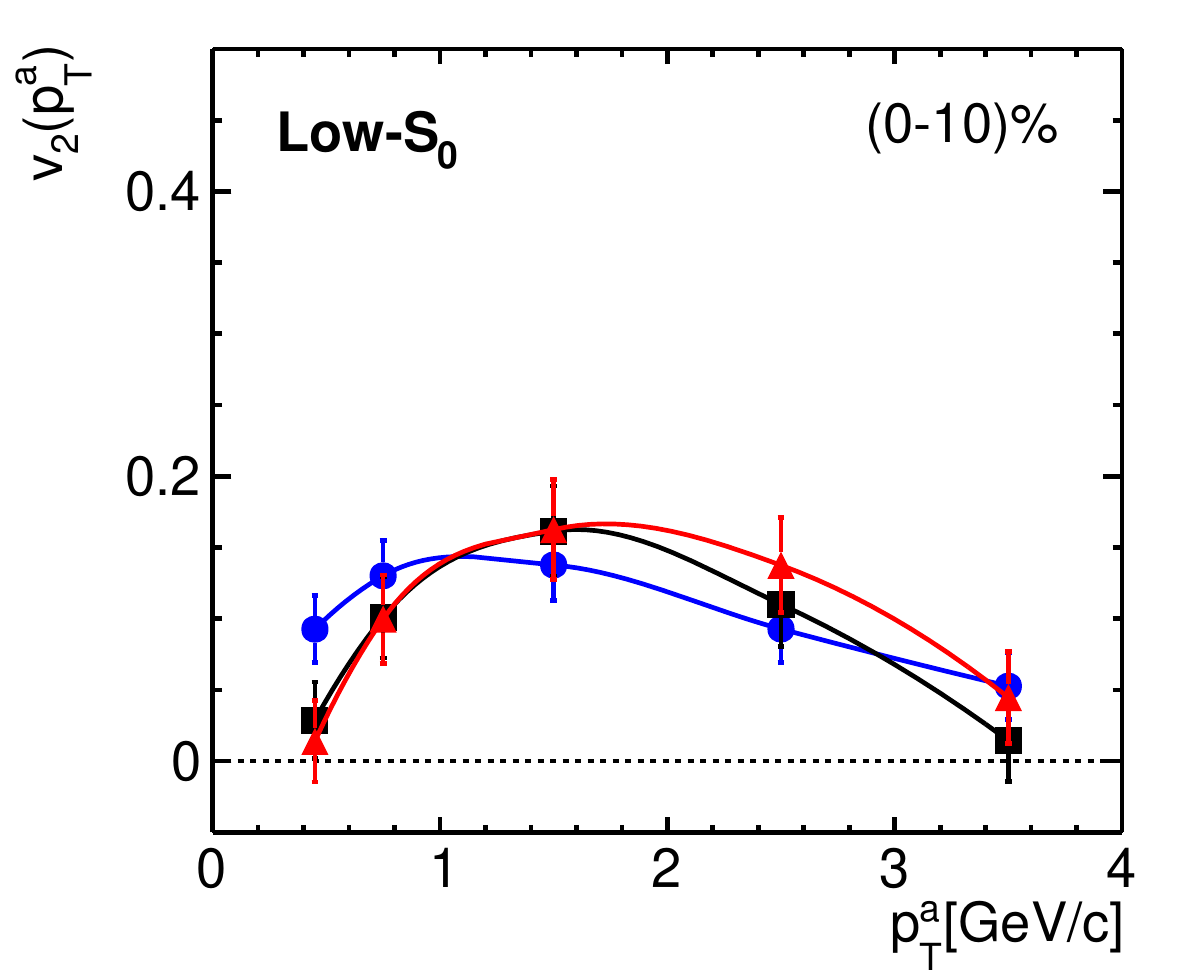}
\includegraphics[scale=0.29]{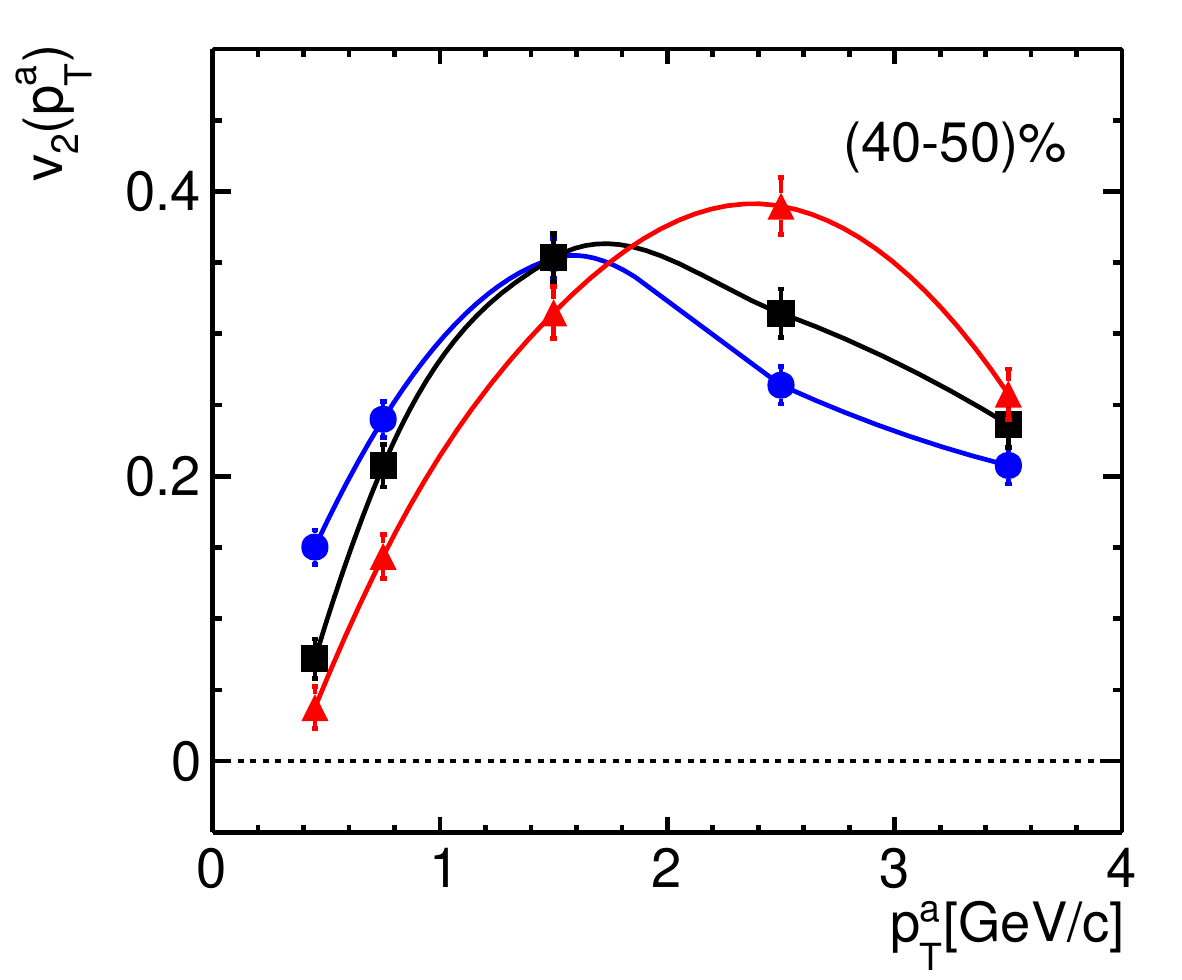}
\includegraphics[scale=0.29]{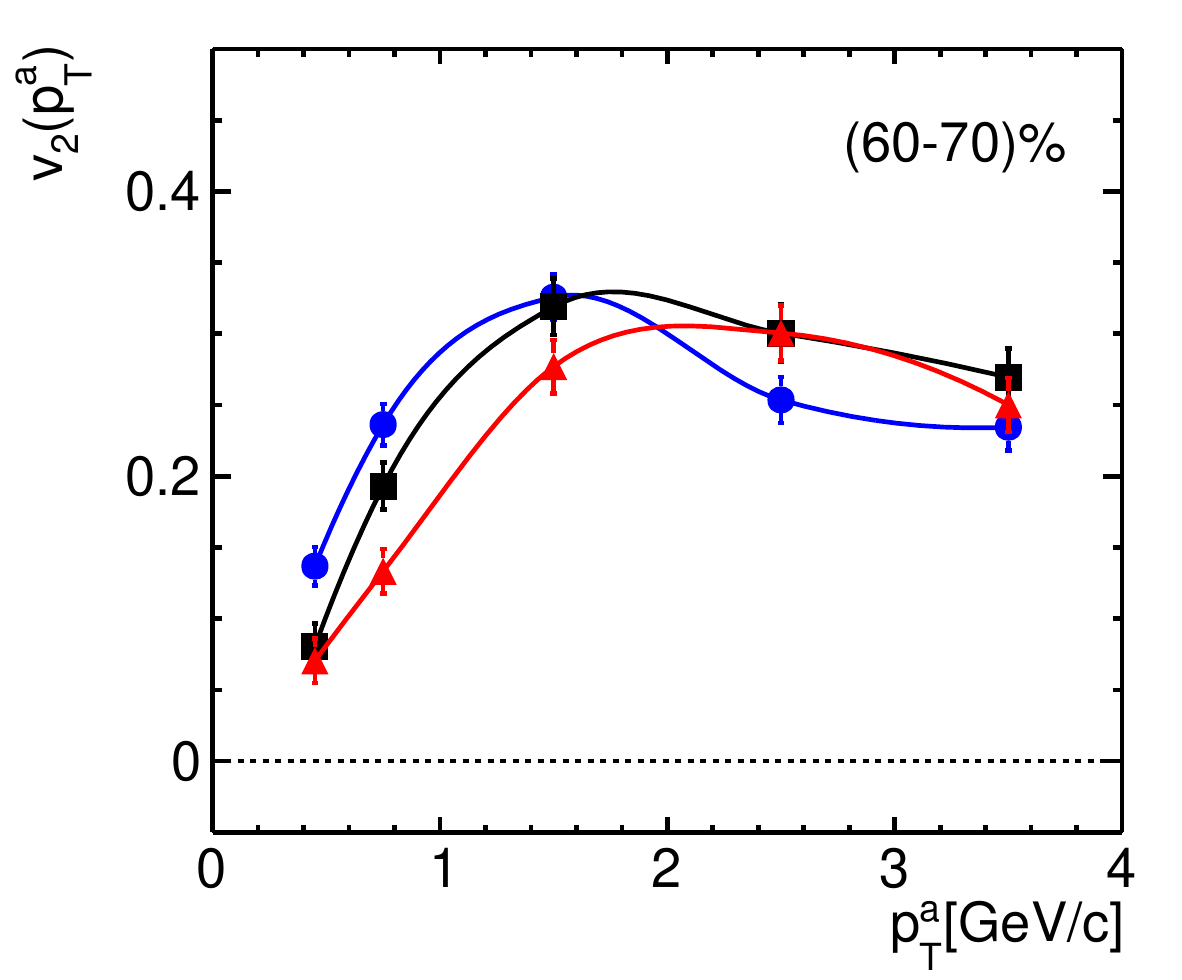}
\caption[width=18cm]{(Color Online) Centrality dependence of elliptic flow co-efficient ($v_{2}(p_{\rm T}^a$)) as a function of $p_{\rm T}^a$ of pions (blue circles), kaons (black squares) and protons (red triangles) for spherocity-integrated (top panel) and low-$S_0$ (bottom panel) events in Pb+Pb collisions at $\sqrt{s_{\rm NN}} = 5.02$ TeV using AMPT model.}
\label{fig6}
\end{center}
\end{figure*}

For $n$ = 2, Eq.~\ref{eq2} gives the second order harmonics in the expansion and its coefficient, $v_2$ provides the numerical measure of the elliptic flow which can be obtained from Eq.~\ref{eq2} using orthogonality relations:
 \begin{eqnarray}
v_{2} = \langle \cos(2(\phi - \psi_2))\rangle
\label{eq3}
% v_{2} = \langle \cos(2\phi)\rangle
\end{eqnarray}
Due to the almond-shaped nuclear overlap region in a non-central heavy-ion collision, the azimuthal anisotropy is expected to be dominated by the $v_2$ component. In addition, the signatures of $v_1$ (directed flow) and $v_3$ (triangular flow) are observed but their contribution is quite less as compared to $v_2$ in a non-central heavy-ion collision. Another interesting observable owing to anisotropic flow and properties of the medium is the correlation function between two particles in relative pseudorapidity ($\Delta\eta = \eta_a - \eta_b$) and azimuthal angle ($\Delta\phi = \phi_a - \phi_b$). The labels $a$ and $b$ denote the two particles in the pair, which are chosen from different (or same) transverse momentum intervals. In this study, we have taken particles in $|\eta|<$ 2.5  to include wider range of particles mostly contributing to $v_2$ and $p_{\rm T}>$ 0.5 GeV/c to be consistent with Refs.~\cite{Aad:2015lwa,Mallick:2021rsd}.
 The two particle correlation function, $C(\Delta\eta ,\Delta\phi)$ can be constructed as the ratio of distributions for same-event pairs, $S(\Delta\eta ,\Delta\phi)$ and mixed-event pairs, $B(\Delta\eta ,\Delta\phi)$, which is given by~\cite{ATLAS:2012at,Aad:2015lwa}:
\begin{eqnarray}
C(\Delta\eta ,\Delta\phi) = \frac{S(\Delta\eta ,\Delta\phi)}{B(\Delta\eta ,\Delta\phi)}
\label{eq4}
\end{eqnarray}
We have chosen five randomly selected events in the same centrality class for the mixed-event pairing and hence the pairs contain no physical correlations whatsoever.
To obtain the anisotropy in the azimuthal direction, our analysis focuses mainly on the shape of the 1D correlation function $\Delta\phi$, given as~\cite{ATLAS:2012at,Aad:2015lwa};
\begin{eqnarray}
C(\Delta\phi)= \frac{dN_{\rm pairs}}{d\Delta\phi} = A \times \frac{\int S(\Delta\eta ,\Delta\phi) d\Delta\eta}{\int B(\Delta\eta ,\Delta\phi)d\Delta\eta}.
\label{eq5}
\end{eqnarray}
Here, the normalization constant $A$ ensures that the number of pairs is the same between the signal ($S$) and background ($B$) events for a given $\Delta\eta$ interval. The $\Delta\eta$ interval is chosen carefully by excluding the jet peak region observed in the $C(\Delta\eta ,\Delta\phi)$ distribution. For our case, the interval is taken to be $2.0< |\Delta\eta| < 4.8$. By choosing particle pairs with a proper pseudorapidity cut, the two-particle correlation method removes the residual non-flow effects in the estimation of elliptic flow. Non-flow azimuthal correlations usually arise from jets and short range resonance decays, which are not associated with the anisotropies in the initial-state collision geometry. Although the non-flow effects for elliptic flow estimation are very small in heavy-ion collisions at very high energies, our study based on event shape dependence being sensitive to jets may get affected by these effects and hence we proceed with this method. To ensure the consistency of the method, one can have a look at Fig. 3 of Ref.~\cite{Mallick:2021rsd}, where the comparison of calculated $v_{2,2}$ between PYTHIA8 (Angantyr)~\cite{Bierlich:2018xfw} and AMPT model is shown.  It is shown that the method removes significant residual non-flow contributions without affecting the event shape studies based on transverse spherocity. In the current study, the non-flow effects and autocorrelation bias are expected to be small but they are still non-zero. Thus, the quantitative interpretation of the results shown in this paper should be taken with caution. A detailed study on the complete removal of non-flow effects and autocorrelation bias is outside the scope of this manuscript.

The pair distribution ($N_{\rm pairs}$) can be expanded in $\Delta\phi$ into a Fourier series~\cite{ATLAS:2012at,Aad:2015lwa}:
\begin{eqnarray}
\frac{dN_{\rm pairs}}{d\Delta\phi} \propto  \bigg[1+2\sum_{n=1}^\infty v_{n,n}(p_{\rm T}^{a},p_{\rm T}^{b}) \cos n\Delta\phi  \bigg].
\label{eq6}
\end{eqnarray}
Here, $v_{n,n}$ is called as the two particle flow coefficient. By rewriting Eq.~\ref{eq6} with the help of Eq.~\ref{eq5} one obtains~\cite{ATLAS:2012at,Aad:2015lwa}:
\begin{eqnarray}
C(\Delta\phi) \propto \bigg[1+2\sum_{n=1}^\infty v_{n,n}(p_{\rm T}^{a},p_{\rm T}^{b}) \cos n\Delta\phi \bigg].
\label{eq7}
\end{eqnarray}

\begin{figure*}[ht!]
\begin{center}
\includegraphics[scale=0.29]{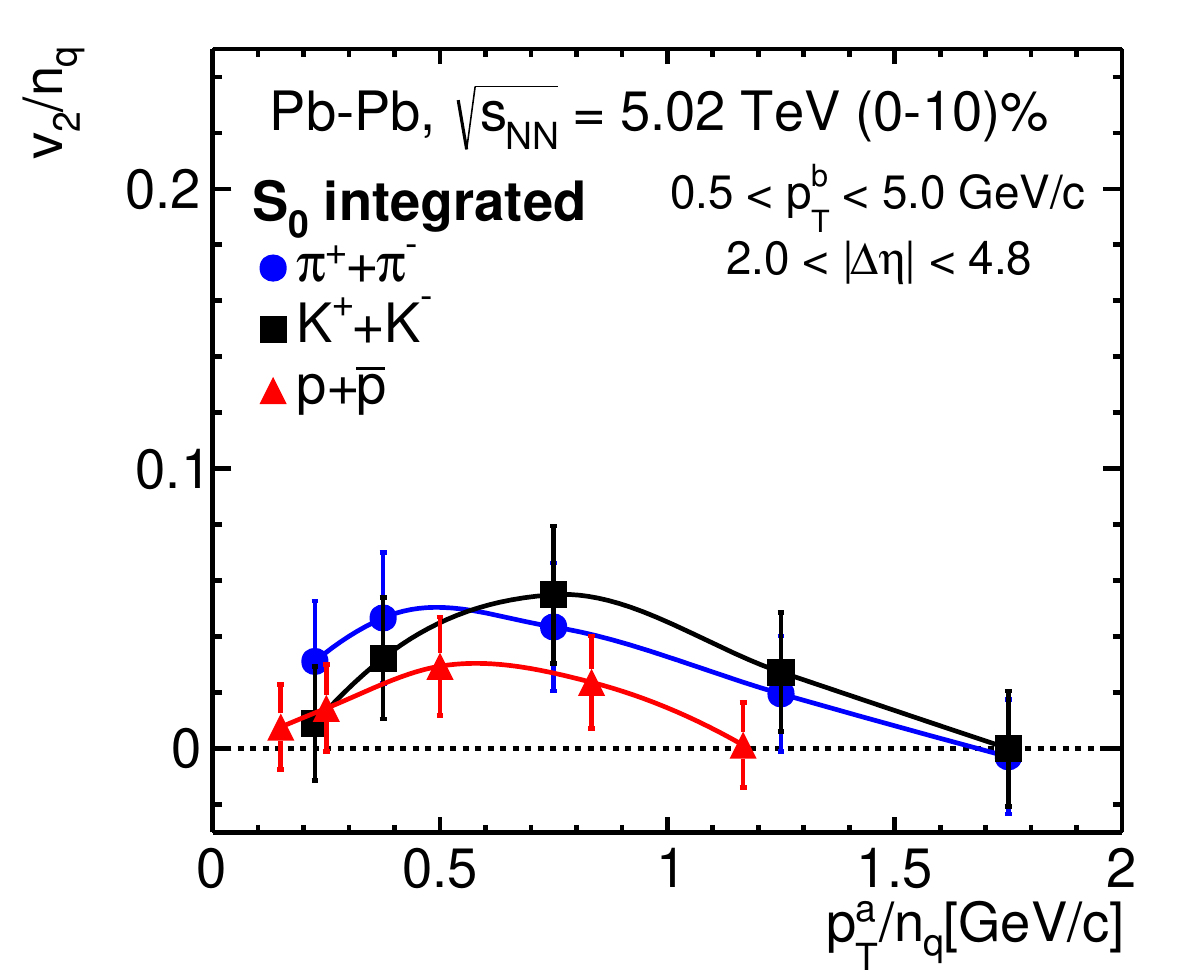}
\includegraphics[scale=0.29]{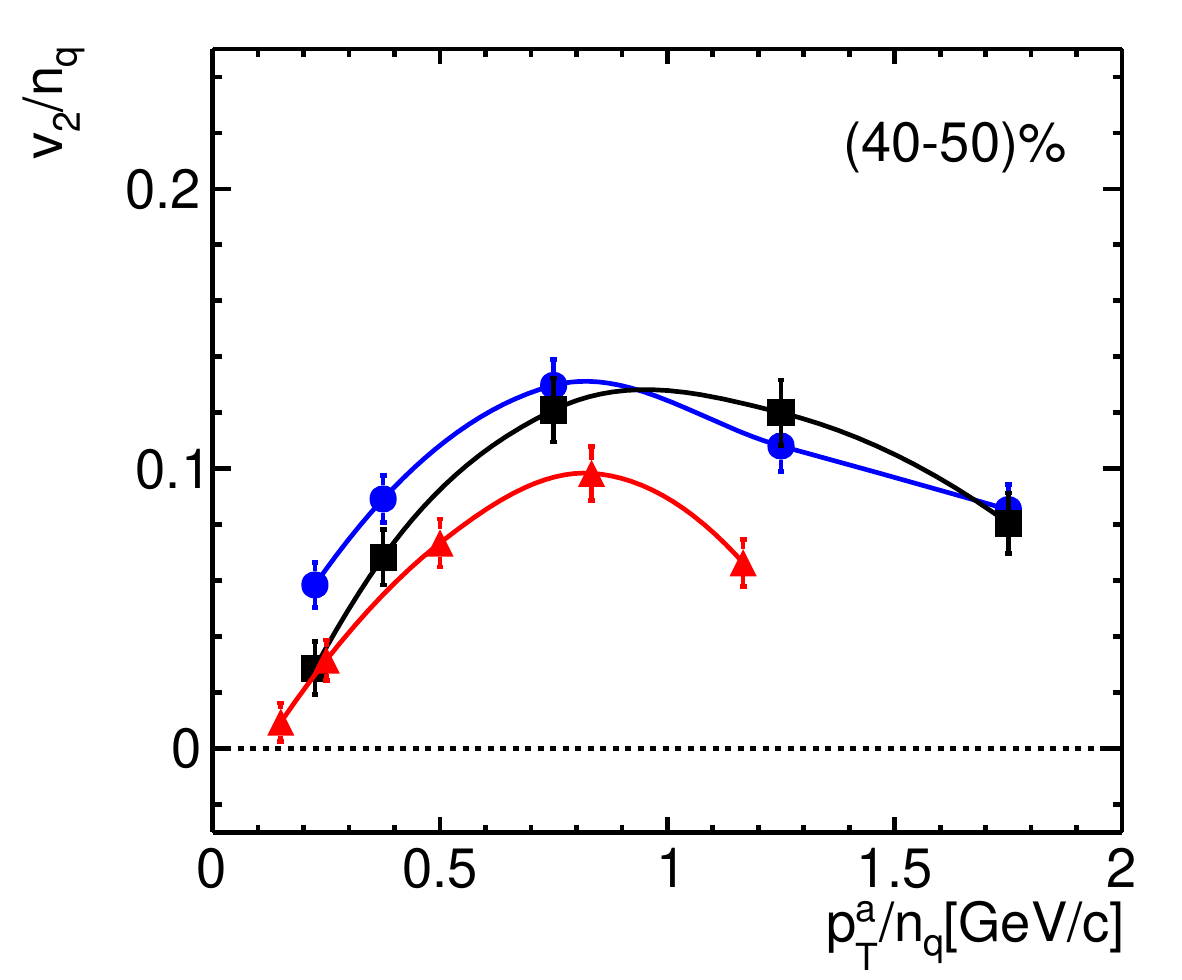}
\includegraphics[scale=0.29]{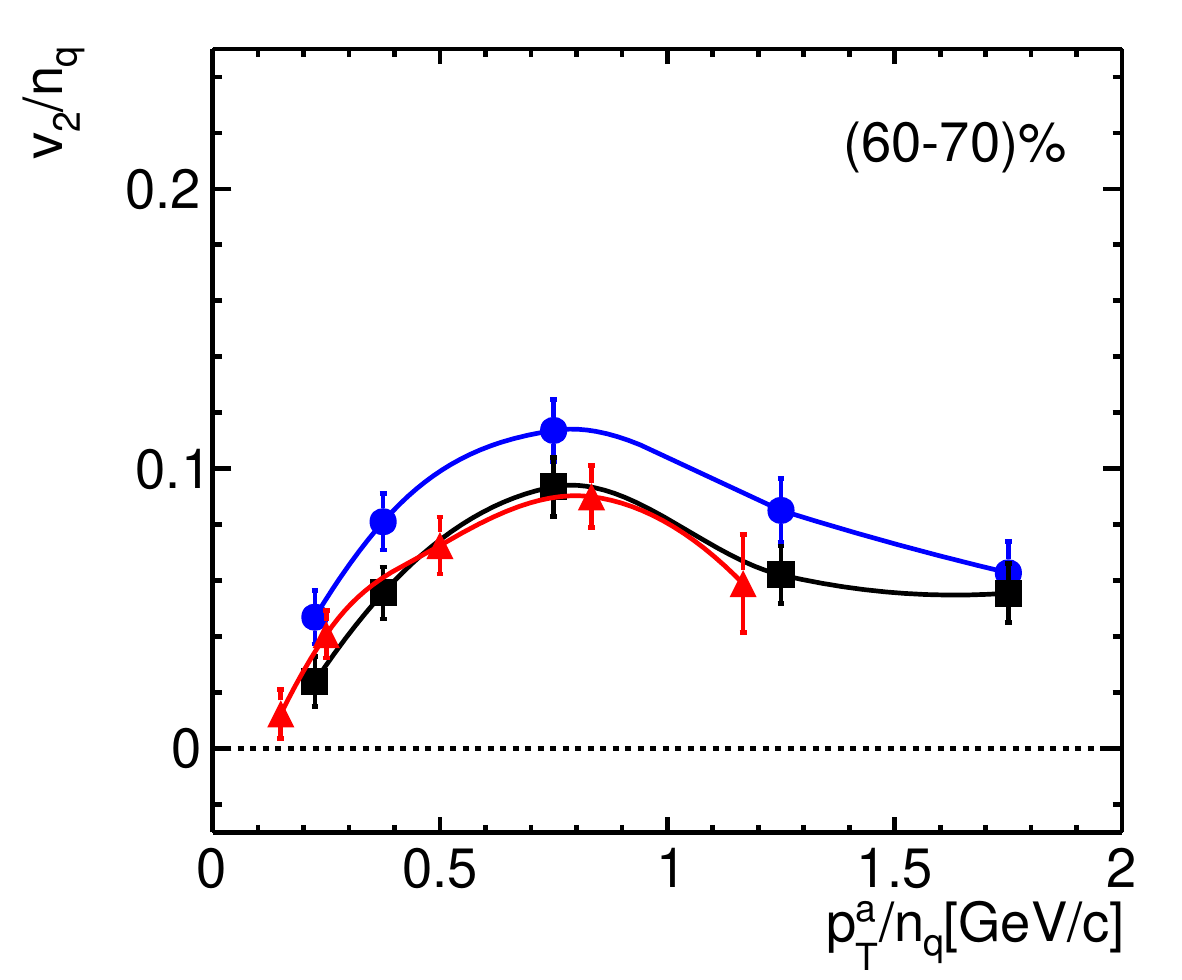}
\includegraphics[scale=0.29]{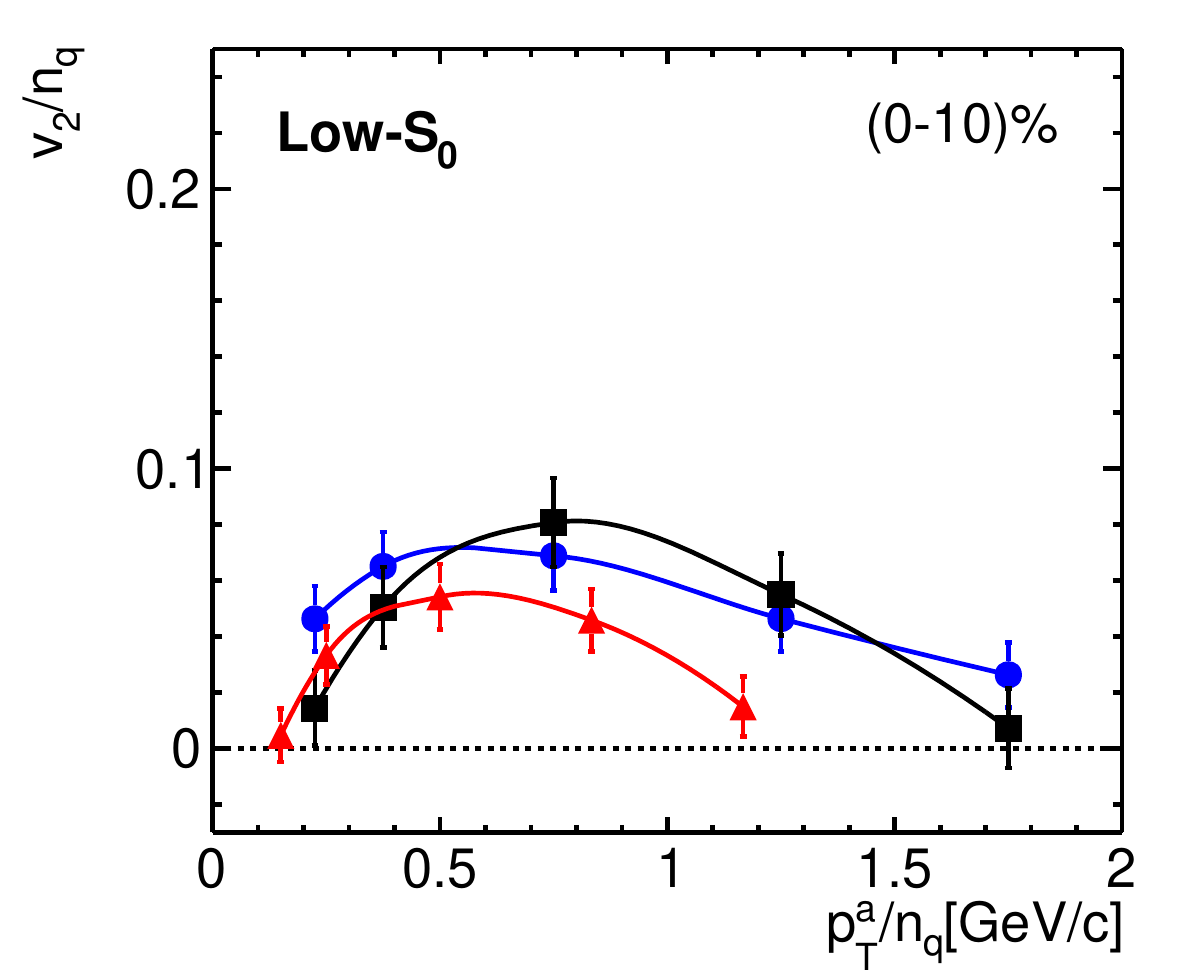}
\includegraphics[scale=0.29]{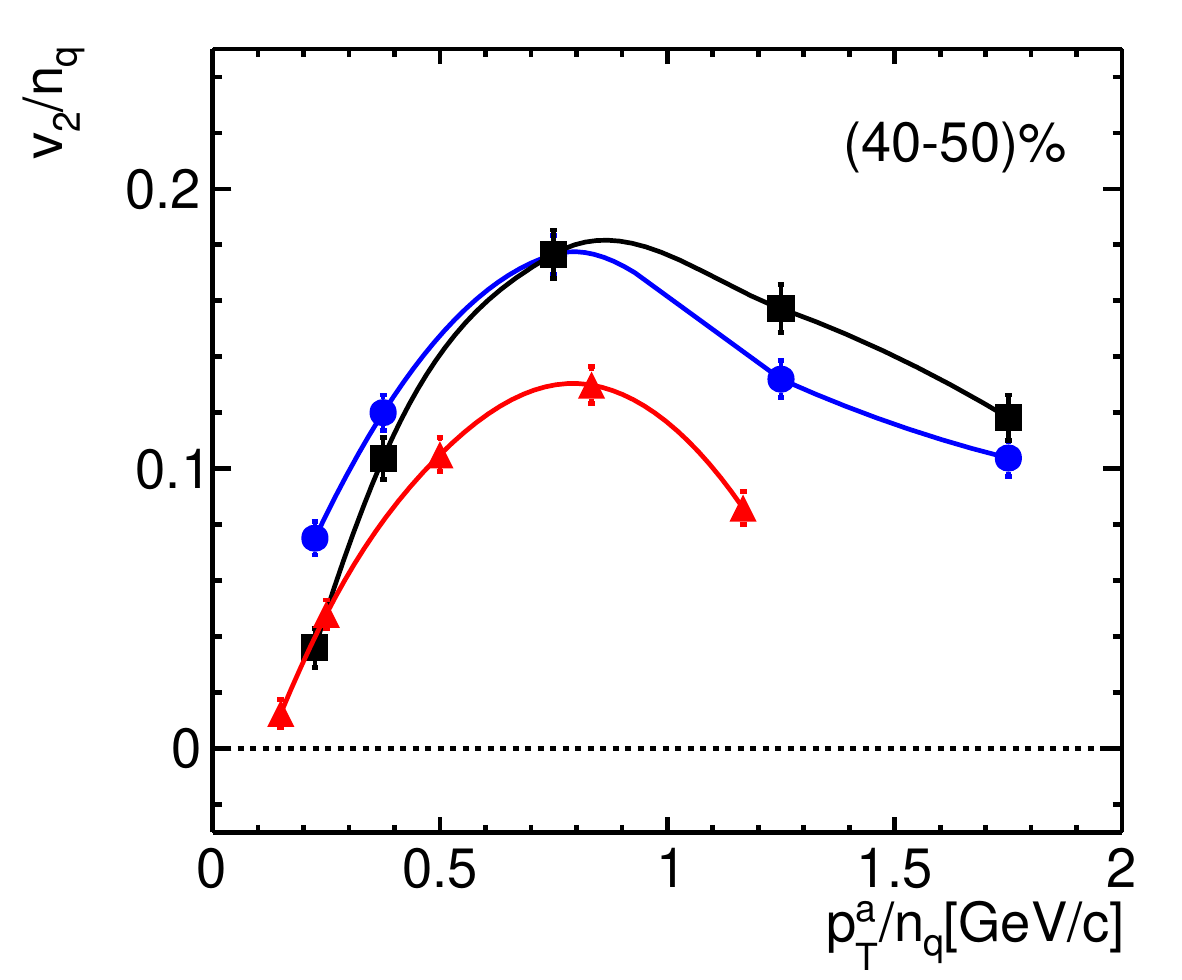}
\includegraphics[scale=0.29]{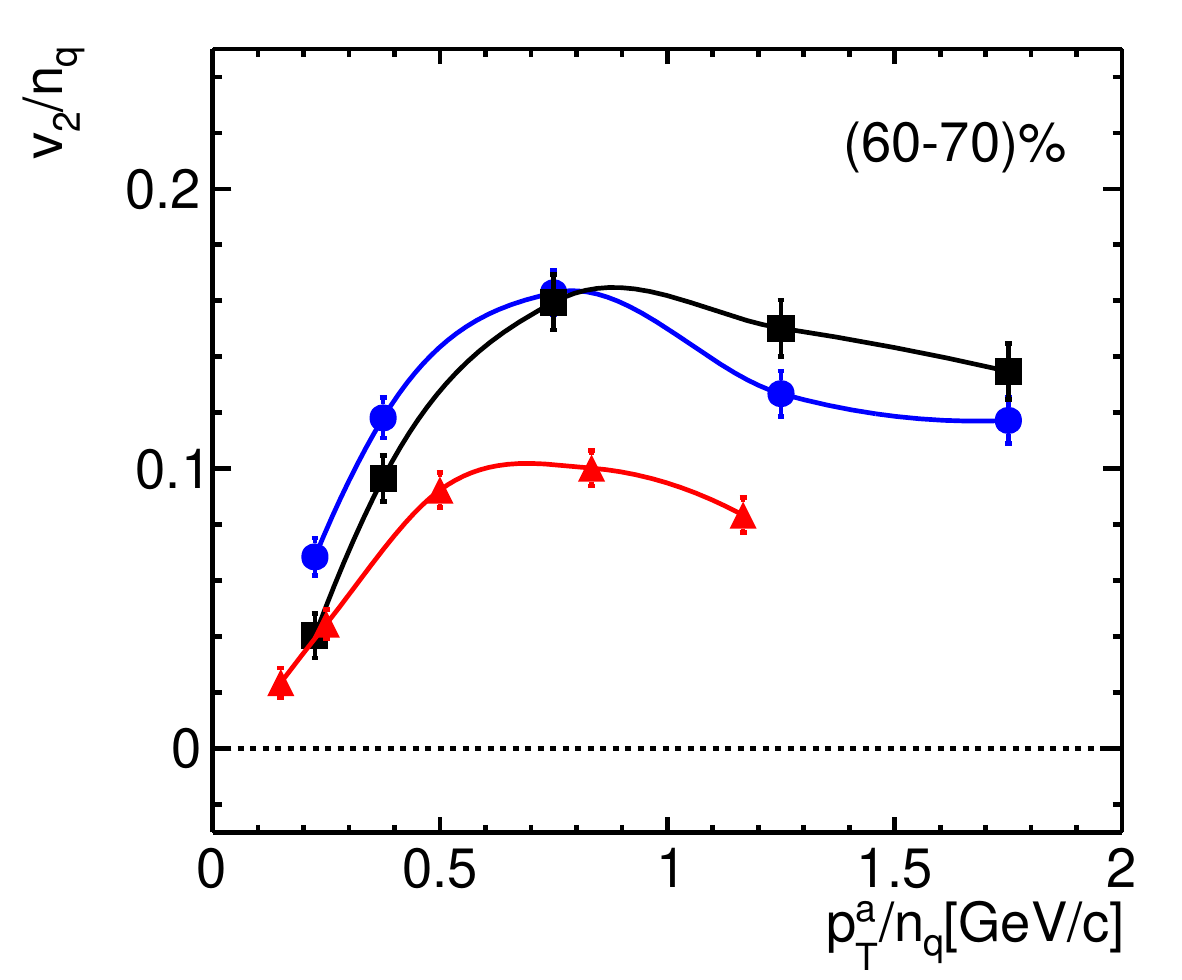}
\caption[width=18cm]{(Color Online) Centrality dependence of elliptic flow co-efficient with number of quark participant scaling ($v_{2}(p_{\rm T}^a$)/$n_q$) as a function of $p_{\rm T}^a$ of pions (blue circles), kaons (black squares) and protons (red triangles) for spherocity-integrated (top panel) and low-$S_0$ (bottom panel) events in Pb+Pb collisions at $\sqrt{s_{\rm NN}} = 5.02$ TeV using AMPT model. }
\label{fig7}
\end{center}
\end{figure*}

Now, $v_{n,n}$ can be calculated as~\cite{ATLAS:2012at,Aad:2015lwa}:
\begin{eqnarray}
v_{n,n}(p_{\rm T}^{a}, p_{\rm T}^{b}) = \langle cos(n\Delta\phi) \rangle
\label{eq8}
\end{eqnarray}
We have taken very narrow bins in $-\pi/2<\Delta\phi<3\pi/2$. $v_{n,n}$ are symmetric functions with respect to $p_{\rm T}^{a}$ and $p_{\rm T}^{b}$. The definition of harmonics defined in Eq.~\ref{eq2} also enters in Eq.~\ref{eq6}, which can be rewritten as~\cite{ATLAS:2012at,Aad:2015lwa};
\begin{eqnarray}
\frac{dN_{\rm pairs}}{d\Delta\phi} \propto  \bigg[ 1+2\sum_{n=1}^\infty v_{n}(p_{\rm T}^{a}) v_{n}(p_{\rm T}^{b}) \cos n\Delta\phi  \bigg].
\label{eq9}
\end{eqnarray}

For simplicity, here we assume no correlation of symmetry plane angles ($\psi_n$) with pseudorapidity, and $\psi_n$ is assumed to be a global phase angle for all particles of the entire event, which is canceled when taking the azimuthal angle difference between two particles. However, one has to note that recent studies suggest pseudorapidity-dependent event plane fluctuations~\cite{CMS:2015xmx,ATLAS:2017rij}.  If the azimuthal anisotropy is driven by collective expansion, then $v_{n,n}$ should factorize into the product of two single-particle harmonic coefficients~\cite{ALICE:2011svq,PHENIX:2002hqx,ATLAS:2012at,CMS:2013wjq}. 
\begin{eqnarray}
v_{n,n}(p_{\rm T}^{a},p_{\rm T}^{b})= v_{n}(p_{\rm T}^{a}) v_{n}(p_{\rm T}^{b}).
\label{eq10}
\end{eqnarray}
From Eq.~\ref{eq10} we derive the single particle flow coefficient as~\cite{ATLAS:2012at,Aad:2015lwa}:
\begin{eqnarray}
v_{n}(p_{\rm T}^{a})= v_{n,n}(p_{\rm T}^{a},p_{\rm T}^{b})/\sqrt{v_{n,n}(p_{\rm T}^{b},p_{\rm T}^{b})}
\label{eq11}
\end{eqnarray}

\begin{figure*}[ht!]
\begin{center}
\includegraphics[scale=0.29]{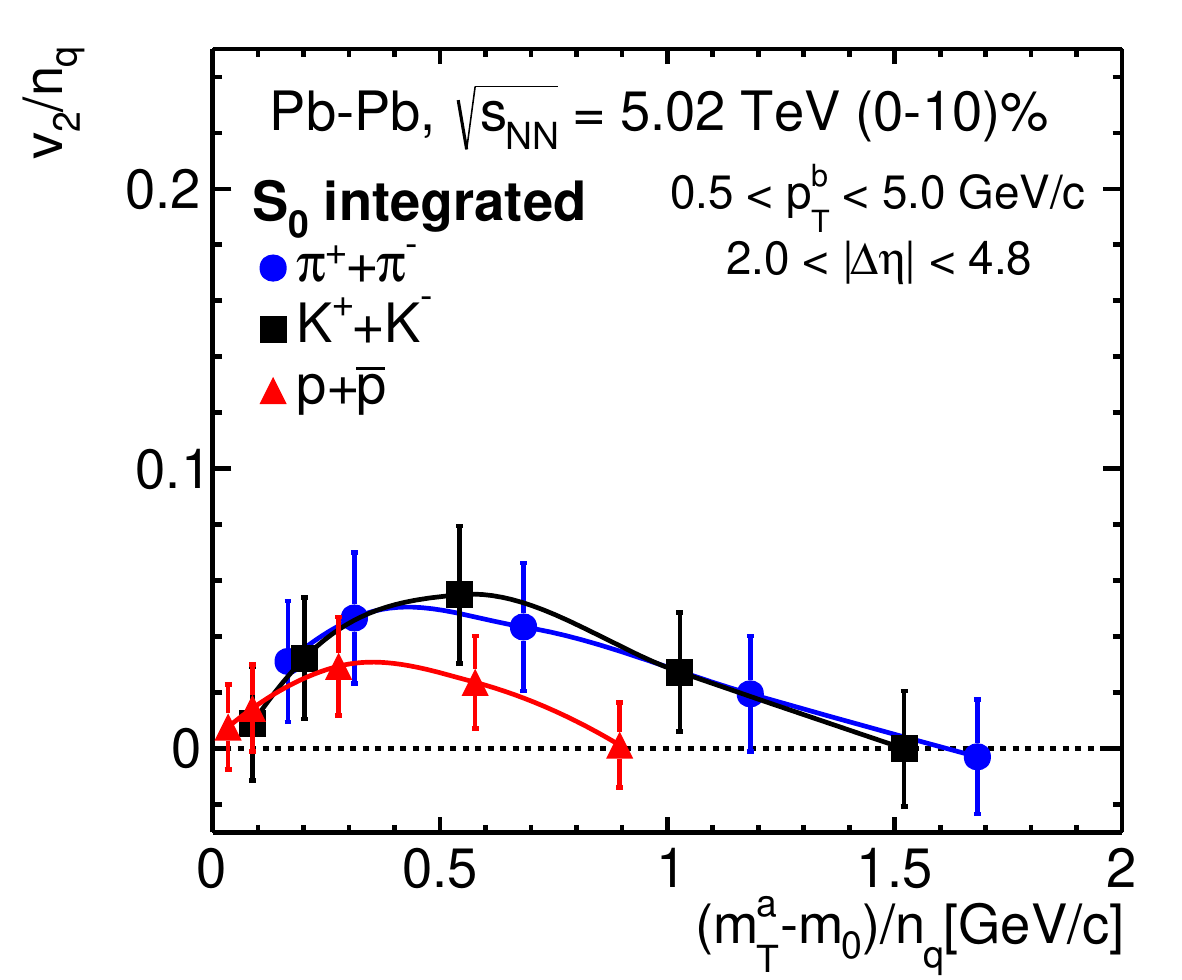}
\includegraphics[scale=0.29]{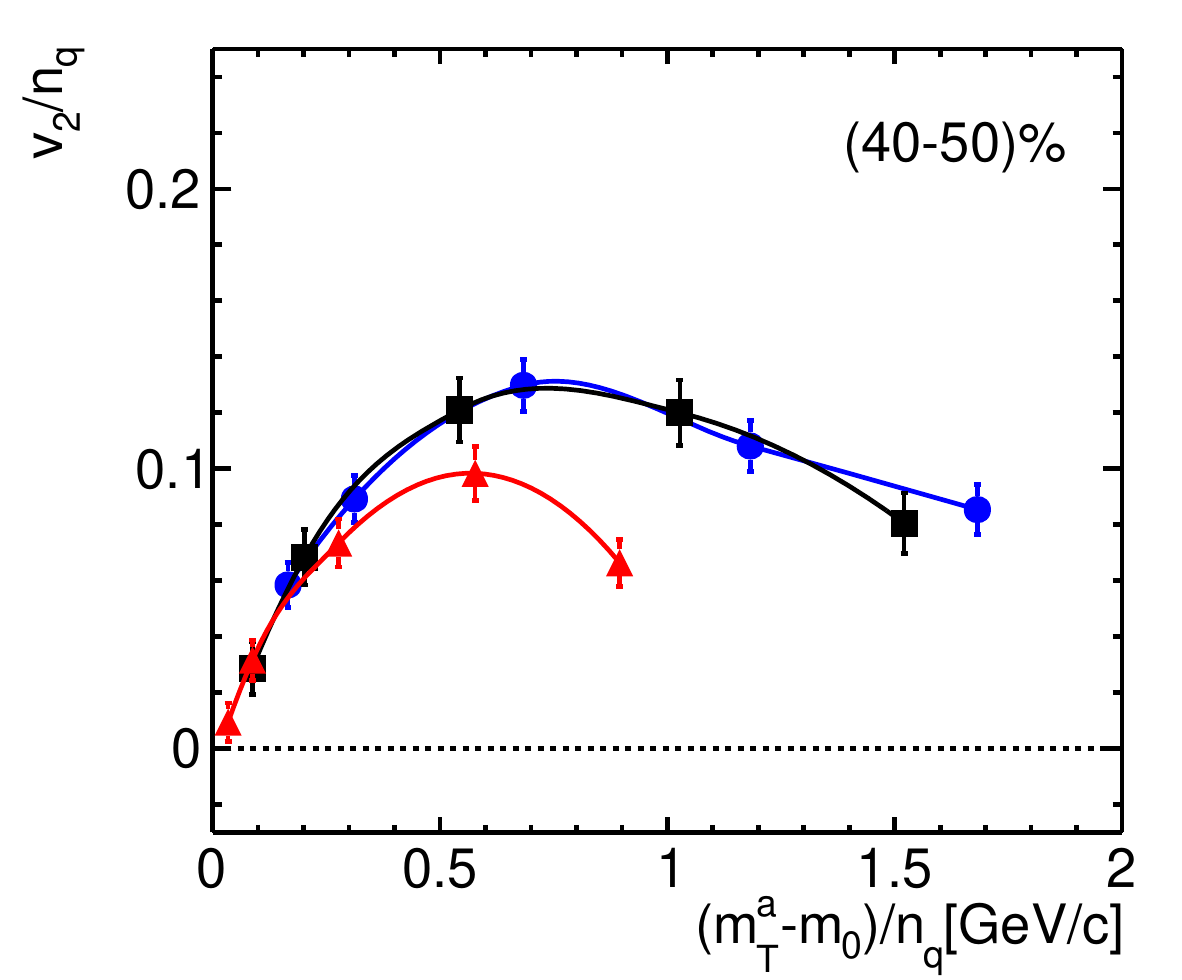}
\includegraphics[scale=0.29]{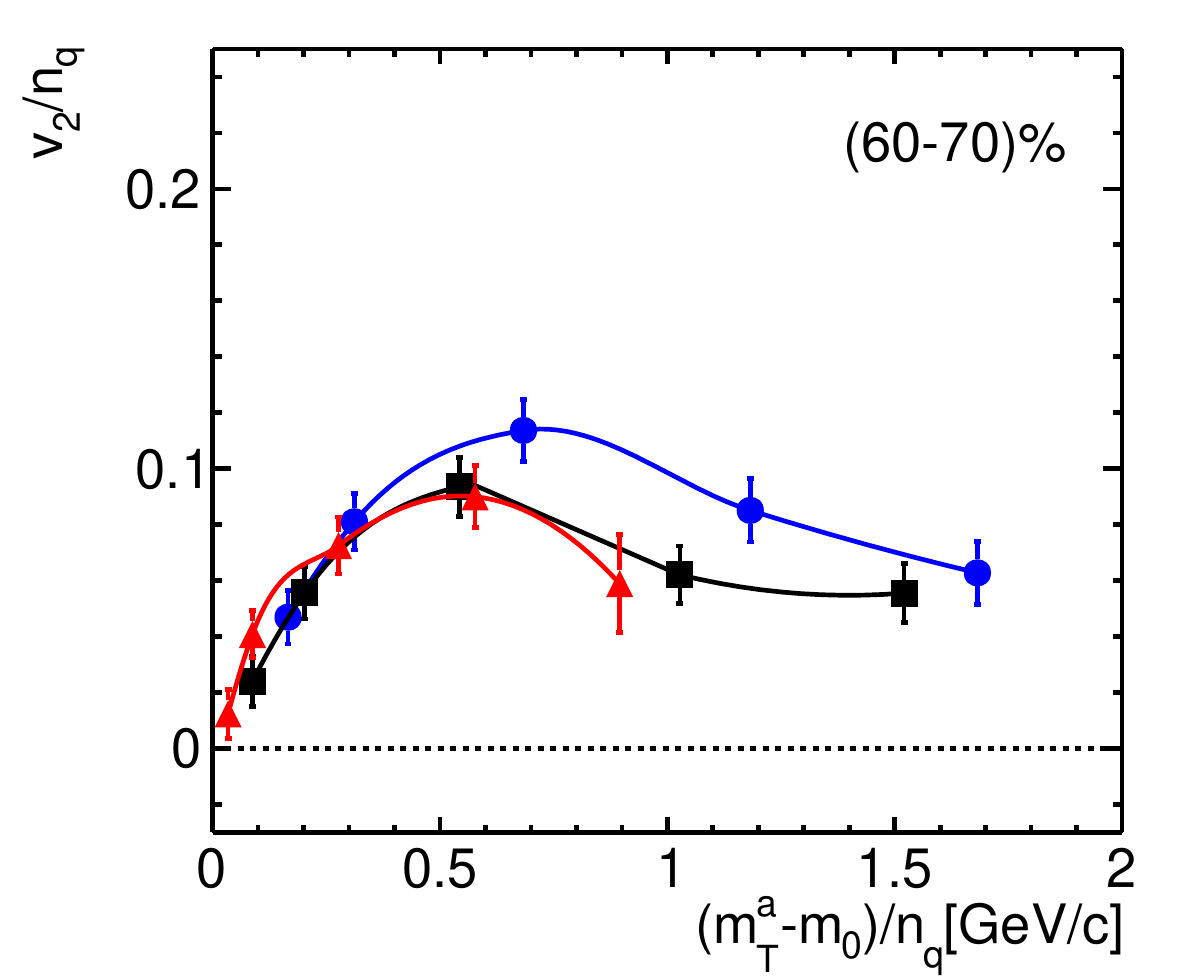}
\includegraphics[scale=0.29]{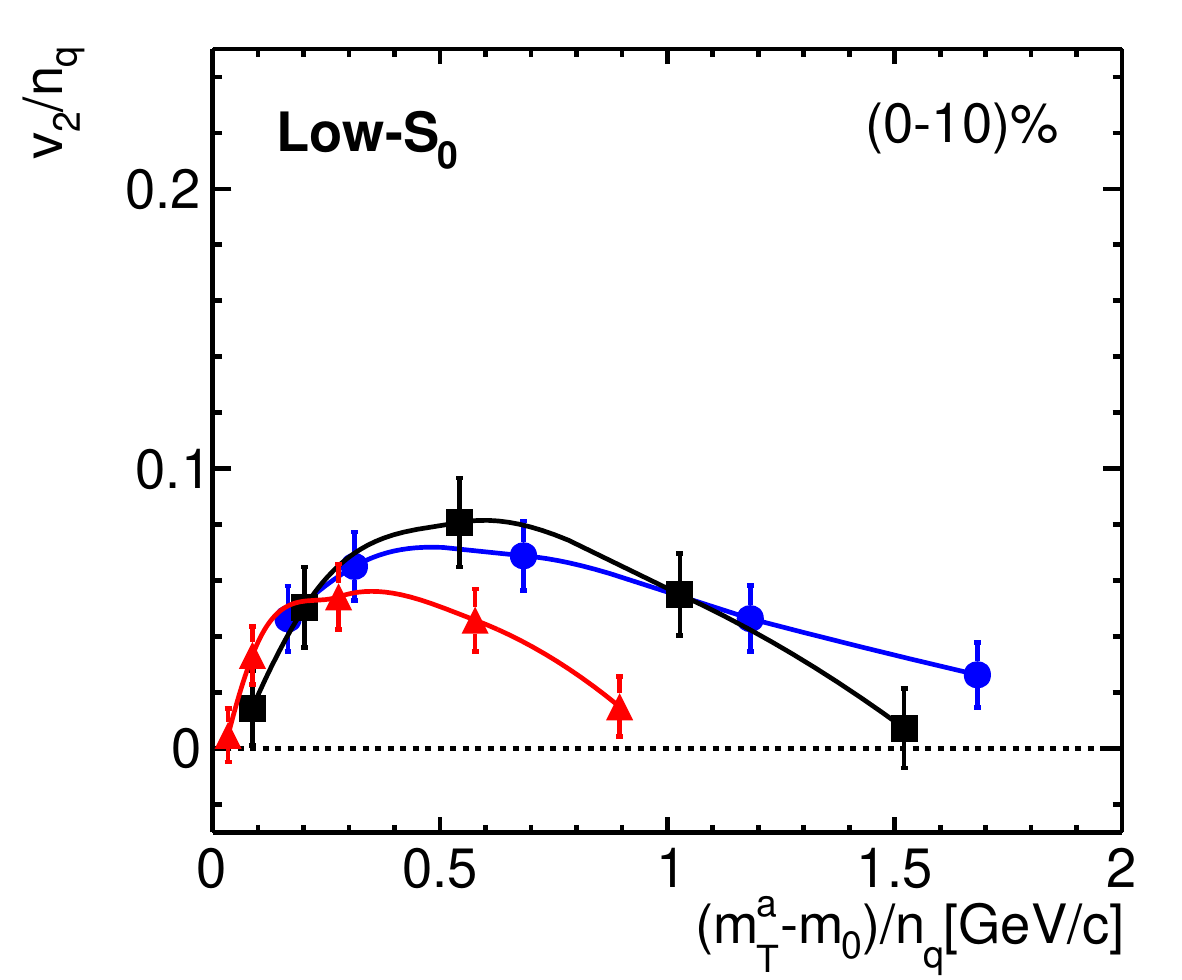}
\includegraphics[scale=0.29]{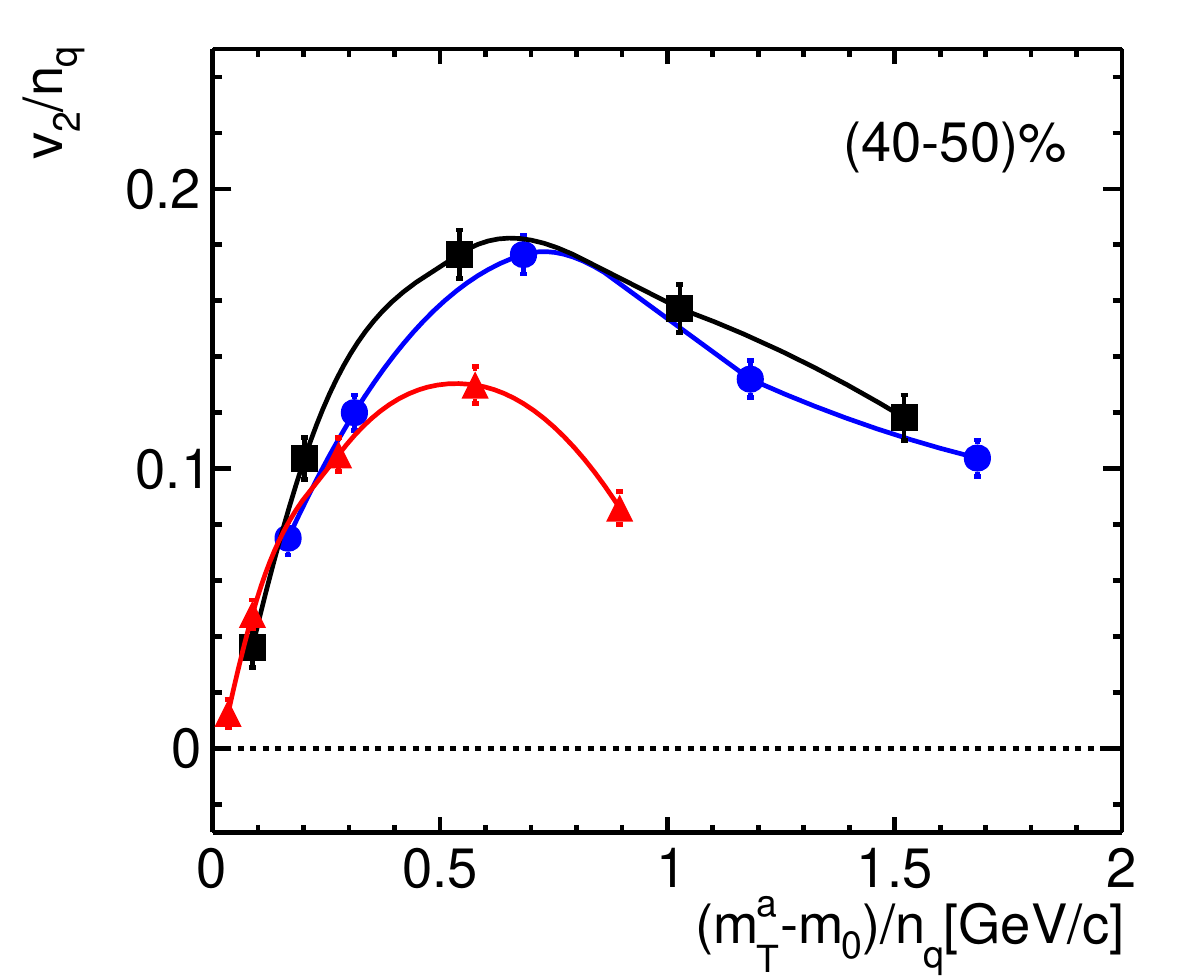}
\includegraphics[scale=0.29]{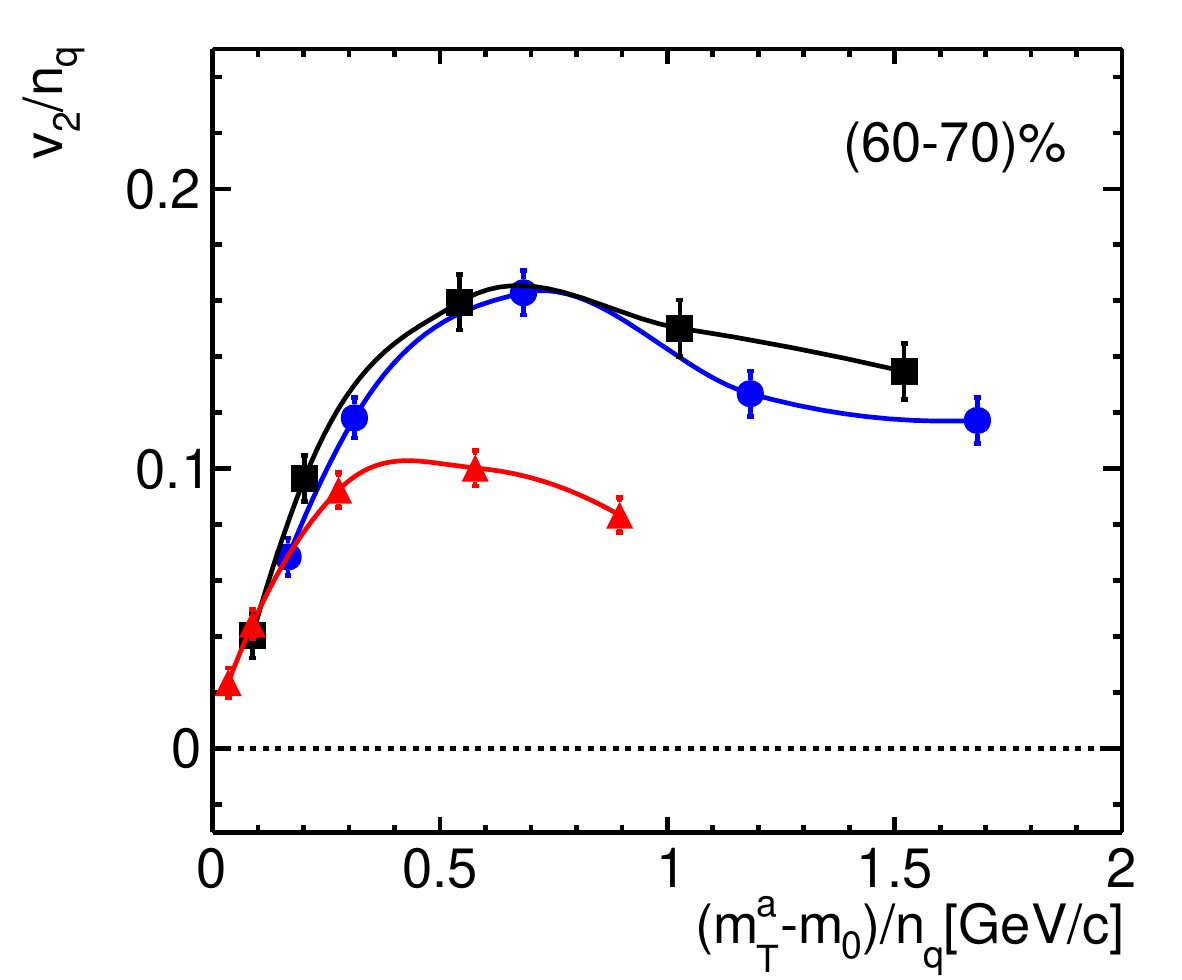}
\caption[width=18cm]{(Color Online) Centrality dependence of elliptic flow co-efficient with transverse mass ($m_{\rm T}$) scaling as a function of $p_{\rm T}^a$ of pions (blue circles), kaons (black squares) and protons (red triangles) for spherocity-integrated (top panel) and low-$S_0$ (bottom panel) events in Pb+Pb collisions at $\sqrt{s_{\rm NN}} = 5.02$ TeV using AMPT model. }
\label{fig8}
\end{center}
\end{figure*}

We now proceed with the estimation of azimuthal anisotropy of $\pi^++\pi^-$, $K^++K^-$ and $p+\bar{p}$ in different spherocity classes for Pb+Pb collisions at the LHC energies from AMPT. For better readability, here onwards we refer to $\pi^++\pi^-$, $K^++K^-$ and $p+\bar{p}$ as pions, kaons, and protons, respectively.

\section{Results and Discussions}
\label{section3}

Top panel of Fig.~\ref{fig2} shows the $p_{\rm T}$-spectra for pions in (0-10)\% Pb+Pb collisions for high-$S_{0}$, $S_0$-integrated and low-$S_{0}$ events and bottom panel shows the ratio of $p_{\rm T}$-spectra for high-$S_{0}$ and low-$S_{0}$ events to the $S_0$-integrated events. In the low-$p_{\rm T}$ region, we observe that pion production is dominated by high-$S_{0}$ events. But as we move on to higher $p_{\rm T}$, the relative yield from high-$S_{0}$ and low-$S_{0}$ events are similar. These results indicate that spherocity can differentiate heavy-ion collisions' event topology based on their geometrical shapes i.e. high-$S_0$ and low-$S_0$. 

Figure~\ref{fig3} shows one dimensional azimuthal correlation of pions for low-$S_0$, high-$S_0$ and spherocity-integrated events in 40-50\% central Pb+Pb collisions at $\sqrt{s_{\rm NN}} = 5.02$ TeV. As described in the previous section, the correlation function is calculated in the $p_{\rm T}$ range of 0.5 to 5 GeV/c for $2.0< |\Delta\eta| < 4.8$. We see a strong correlation of modulation magnitude with spherocity in Fig.~\ref{fig3}. Also, a double peak is observed for the away side ($\Delta\phi \sim \pi$) for high-$S_0$ events, which may arise from the residual contribution of the triangular flow. The observed behavior is analogous to the behavior observed in Refs.~\cite{Aad:2015lwa}, where the flow vector is used as an event classifier for heavy-ion collisions. Figures~\ref{fig4} and~\ref{fig5} show the two-particle and single-particle elliptic flow coefficients of pions for low-$S_0$, high-$S_0$ and spherocity-integrated events in 40-50\% central Pb+Pb collisions at $\sqrt{s_{\rm NN}} = 5.02$ TeV obtained using Eqs.~\ref{eq8} and ~\ref{eq11}, respectively. We observe that the contribution towards elliptic flow coefficients is dominated by low-$S_0$ events while high-$S_0$ events show nearly zero elliptic flow coefficients. In a similar fashion, we obtain the elliptic flow coefficients for pions, kaons, and protons and then study the number of constituent quark scaling (NCQ) of the elliptic flow coefficients.

Figure~\ref{fig6} shows the single particle elliptic flow co-efficient ($v_{2}(p_{\rm T}^a$)) as a function of $p_{\rm T}^a$ for pions, kaons and protons in (0-10)\%, (40-50)\% and (60-70)\% central Pb+Pb collisions at $\sqrt{s_{\rm NN}} = 5.02$ TeV for spherocity-integrated and low-$S_0$ classes. The values of $v_{2}(p_{\rm T}^a$) increases from (0-10)\% to (40-50)\% central collisions while slightly decreases for 
(60-70)\% central collisions for all particle species. The increasing behavior is consistent with the picture of the geometry of the collision driving the final state anisotropy. The slight decrease for (60-70)\% central collisions might have originated from an interplay of different effects like the smaller lifetime of the fireball in peripheral collisions with respect to most central collisions which does not allow the elliptic flow to further develop and less significant contribution of eccentricity fluctuations and to final state hadronic effects~\cite{Song:2013qma}. For (40-50)\% and (60-70)\% central collisions, we see a clear mass hierarchy in the low-$p_{\rm T}$ region ($p_{\rm T}<$~2 GeV/c) and in the intermediate-$p_{\rm T}$ the elliptic flow for protons are higher than that of pions and kaons. As per hydrodynamic calculations, the mass hierarchy in the low-$p_{\rm T}$ could be an interplay between radial and elliptic flow. The baryon and meson separation in the intermediate-$p_{\rm T}$ could arise from the quark coalescence mechanism of AMPT string melting mode. As described earlier, in the string melting mode of AMPT, when a quark and an anti-quark are close in phase space with their momenta very close to each other, they coalesce to form a meson. When three quarks come closer in phase space, they combine to form a baryon. Thus the probability of recombination of three quarks becomes higher in the intermediate-$p_{\rm T}$ range. 

Owing to the quark coalescence mechanism, one may expect elliptic flow to follow the constituent quark number (NCQ) scaling behavior. If elliptic flow follows NCQ-scaling, then it can be expressed as,
\begin{eqnarray}
v_{2}^{h}(p_{\rm T}^{a}) = n_{q} \times v_{2}^{q}(p_{\rm T}^{a}/n_{q}). 
\label{eq12}
\end{eqnarray}
Here, $n_{q}$ is the number of constituent quarks for h-hadron. In RHIC experimental observations~\cite{Adams:2003am,Abelev:2007qg,Adler:2003kt}, it was found that the identified particles follow the NCQ scaling behavior. The PHENIX collaboration extends the scaling property to the lower transverse momentum by obtaining elliptic flow as a function of the transverse kinetic energy, where the kinetic energy is defined by $m_{\rm T} - m_0$~\cite{Adler:2003kt,Afanasiev:2007tv,Adare:2006ti}. Here, $m_{\rm T}$ is the transverse mass ($\sqrt{p_{\rm T}^2+m_0^2}$) and $m_0$ is the rest mass of the particle. The kinetic energy representation of scaling was found to be valid in the RHIC results. Also, this scaling behavior was successfully reproduced by models which invoke quark coalescence as the dominant hadronization mechanism~\cite{Molnar:2003ff,Greco:2003mm,Fries:2003kq,Hwa:2003ic}. Thus, the NCQ scaling of the elliptic flow has been interpreted as evidence of the dominance of quark degrees of freedom in the initial stages of heavy-ion collisions~\cite{Molnar:2003ff,Greco:2003mm,Fries:2003kq,Hwa:2003ic}. In this work, we have studied both the NCQ-scaling behavior for pions, kaons, and protons. Figure~\ref{fig7} shows centrality dependence of elliptic flow co-efficient with number of quark participant scaling ($v_{2}(p_{\rm T}^a$)/$n_q$) as a function of $p_{\rm T}^a/n_q$ of pions, kaons and protons for spherocity-integrated and low-$S_0$ events.  Figure~\ref{fig8} shows $v_{2}(p_{\rm T}^a$)/$n_q$ scaling as a function of $(m_{\rm T} - m_0)/n_q$. In the intermediate transverse momentum range, where the quark coalescence mechanism is dominant, both the scaling of $v_{2}(p_{\rm T}^a$)/$n_q$ show a significant deviation. Results from LHC suggests that both NCQ and transverse kinetic energy scaling of $v_{2}(p_{\rm T}^a$)/$n_q$ at the LHC energies show a deviation of 20\%~\cite{Abelev:2014pua, v2_ALICE2} for intermediate/high transverse momenta. The observed deviations seem to be similar for all centrality classes.

Let us now discuss how the results vary when such a study is performed in different spherocity classes. In Fig.~\ref{fig6}, for (0-10)\% centrality we see that the values of $v_2$ is similar in both spherocity-integrated and low-$S_0$ events. For (40-50)\% and (60-70)\% centrality classes we see significantly higher $v_2$ in low-$S_0$ events compared to the spherocity-integrated events for all particles. For (40-50)\% centrality class, the baryon-meson separation in intermediate-$p_{\rm T}$ is more prominent in low-$S_0$ events compared to spherocity-integrated events. While studying the NCQ scaling and transverse kinetic energy scaling of the elliptic flow for different spherocity classes in Figs.~\ref{fig7} and~\ref{fig8}, it was found that the low-$S_0$ events show a larger deviation compared to spherocity-integrated events in all centrality classes. The deviation is highly significant, where one can draw a conclusion of violation in the NCQ scaling of elliptic flow at the LHC energies in the AMPT model. This essentially indicates that by using spherocity not only one can study the events with the enhanced elliptic flow but also one can directly probe into the scaling properties of the elliptic flow. 
%new

\begin{figure*}[ht!]
\begin{center}
\includegraphics[scale=0.25]{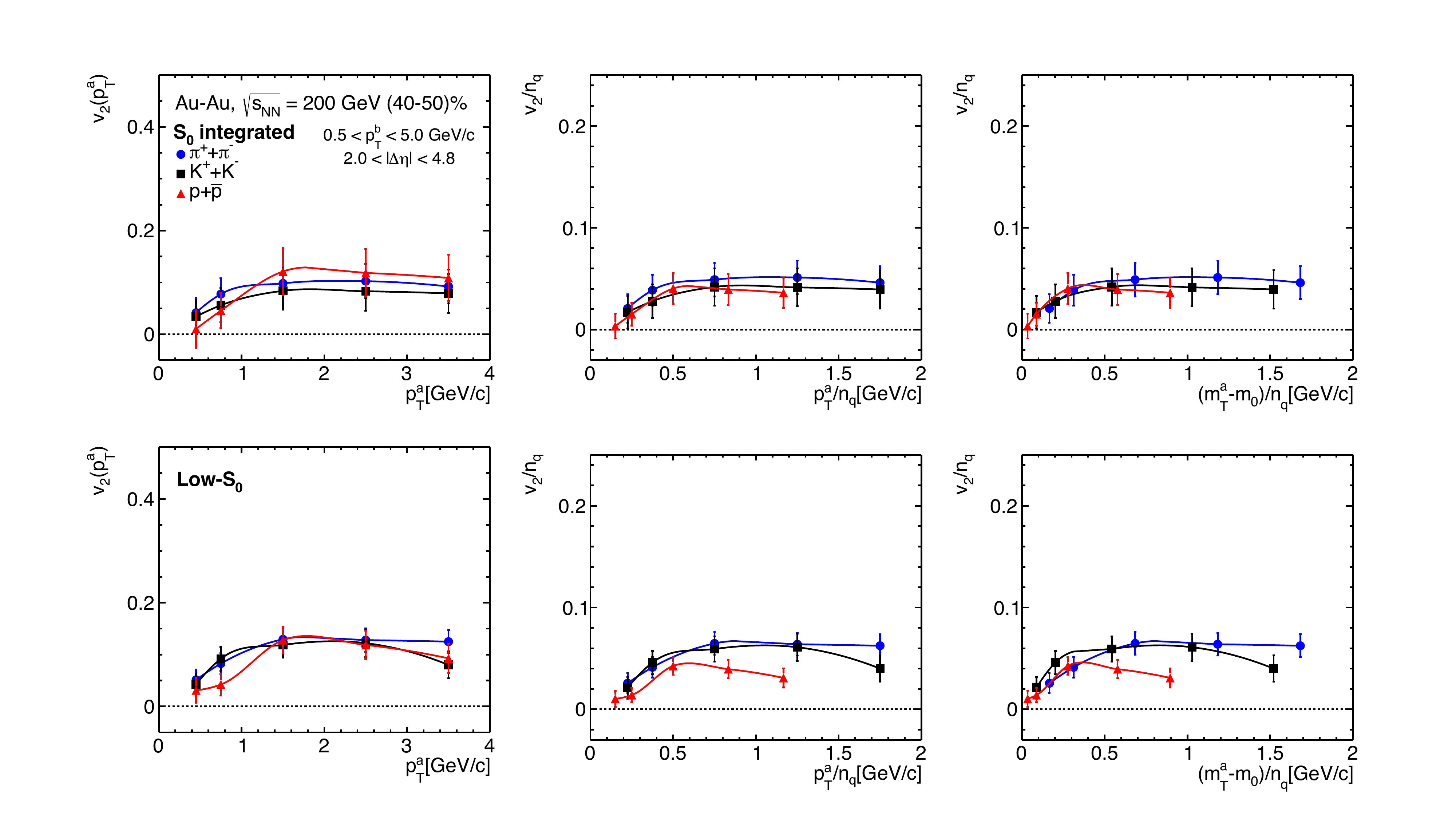}
\caption[width=18cm]{(Color Online) Elliptic flow co-efficient (left) with number of quark participant scaling ($v_{2}(p_{\rm T}^a$)/$n_q$) (middle) and with transverse mass ($m_{\rm T}$) scaling (right) as a function of $p_{\rm T}^a$ of pions (blue circles), kaons (black squares) and protons (red triangles) for spherocity-integrated (top panel) and low-$S_0$ (bottom panel) events in (40-50)\% central Au+Au collisions at $\sqrt{s_{\rm NN}} = 200$ GeV using AMPT model. }
\label{fig9}
\end{center}
\end{figure*}

To strengthen the above conclusion, we have performed a spherocity dependent study on the scaling behavior of $v_2$ in (40-50)\% central Au+Au collisions at $\sqrt{s_{\rm NN}} = 200$ GeV using AMPT model. This study is very important due to the fact that the kinetic energy representation of the scaling of $v_2$ was found to be valid in the RHIC results~\cite{Adler:2003kt,Afanasiev:2007tv,Adare:2006ti}. The upper panels of Fig.~\ref{fig9} show elliptic flow coefficient with the number of quark participant scaling and with transverse mass scaling for spherocity-integrated class. As expected, both the scaling behavior are found to be valid for spherocity-integrated events. However, for low-$S_0$ events, shown in lower panels of Fig.~\ref{fig9}, both the scaling behavior are found to be violated. This study establishes the fact that spherocity can separate out the events where the initial stage of the systems is dominated by the quark degrees of freedom from total events and directly probe into the scaling properties of the elliptic flow. Such observation also strengthens our previous publications using simulation from AMPT model~\cite{Mallick:2021rsd,Mallick:2021wop}, where it was argued that the transverse spherocity can be used as an event classifier in experimental analyses for heavy-ion collisions at RHIC and the LHC.

\section{Summary}
\label{section4}

In summary,

\begin{itemize}
\item The transverse momentum space correlation as a function of transverse spherocity indicates a strong anti-correlation of transverse spherocity with the azimuthal anisotropy in the produced system in heavy-ion collisions. This correlation is further confirmed with the study of transverse spherocity dependent one dimensional azimuthal correlation.
\item The contribution towards elliptic flow coefficients is dominated by low-$S_0$ events while high-$S_0$ events show nearly zero elliptic flow.
\item The baryon-meson separation at the intermediate transverse momentum, the NCQ scaling, and the transverse kinetic energy scaling of $v_{2}(p_{\rm T}^a$)/$n_q$ from AMPT show similar behavior for pions, kaons and protons as observed at the LHC.
\item The baryon-meson separation in intermediate-$p_{\rm T}$ was found to be more prominent in low-$S_0$ events compared to spherocity-integrated events.
\item The study on the NCQ scaling and the transverse kinetic energy scaling of the elliptic flow in different spherocity classes shows that the low-$S_0$ events violate the scaling properties of the elliptic flow at both RHIC and the LHC energies. This essentially indicates that by using spherocity not only one can study the events with the enhanced elliptic flow but also one can directly probe into the scaling properties of the elliptic flow. 
\item The violation of the scaling properties of the elliptic flow at both RHIC and the LHC energies for the low-$S_0$ events
(jetty-like) is in line with the fragmentation process of hadronization expected for the high-momentum partons in jetty-like events.

\end{itemize}

One should note that these results are obtained from an event generator (AMPT) which successfully describes the elliptic flow for spherocity-integrated events in heavy-ion collisions. Thus it is expected that the observed behaviors to mimic the experimental data. However, any deviation of experimental results from AMPT results on elliptic flow could be very interesting to understand the system dynamics. We believe that in the coming days of Run-3 at the LHC, transverse spherocity can serve as an important event classifier that could give a better insight into the possible QGP formation in heavy-ion collisions. Further, it would be of special interest to look into the correlation of spherocity on higher-order harmonics of azimuthal anisotropy.

\section*{Acknowledgements}
 R.S. acknowledges the financial support under the CERN Scientific Associateship and the financial grants under DAE-BRNS Project No. 58/14/29/2019-BRNS of the Government of India. S.T. acknowledges the support under the INFN postdoctoral fellowship. 
 The authors would like to acknowledge the usage of resources of the LHC grid Tier-3 computing facility at IIT Indore.

%\clearpage

\end{document}